\documentclass[lettersize,12pt,onecolumn]{IEEEtran}
\usepackage[dvipsnames]{xcolor}
\usepackage{amsmath,amssymb,amsfonts}
\usepackage[ruled, linesnumbered]{algorithm2e}
\usepackage{algorithmic}
\usepackage{graphicx,color}
\usepackage{textcomp}
\usepackage{url}
\usepackage[hidelinks]{hyperref}
\hypersetup{colorlinks,linkcolor={blue},citecolor={blue},urlcolor={black}}
\usepackage[square,sort,comma,numbers,compress]{natbib}
\usepackage{nicefrac}     
\usepackage{orcidlink}
\usepackage[super]{nth}
\usepackage{soul}
\usepackage{tikz}
\usetikzlibrary{shapes.geometric, arrows.meta, positioning, backgrounds}
\usepackage{setspace}
\usepackage[left=2.5cm, right=2.5cm, top=1.5cm, bottom=1.5cm]{geometry}
\tikzstyle{startstop} = [ellipse, minimum width=2cm, minimum height=0.6cm, text centered, draw=black, fill=red!30]
\tikzstyle{process} = [rectangle, rounded corners, minimum width=2cm, minimum height=0.6cm, text centered, draw=black, fill=blue!10]
\tikzstyle{arrow} = [thick,->,>=stealth]
\tikzstyle{section} = [rectangle, minimum width=2cm, minimum height=0.6cm, text centered, draw=black, fill=gray!20]

\def\bibfont{\footnotesize} 
\def\BibTeX{{\rm B\kern-.05em{\sc i\kern-.025em b}\kern-.08em
    T\kern-.1667em\lower.7ex\hbox{E}\kern-.125emX}}
    
\begin{document}

\title{Integrated Access and Backhaul (IAB) in Low Altitude Platforms}
\author{\IEEEauthorblockN{Reza Ghasemi Alavicheh\textsuperscript{\textsf{\scalebox{1.3}{\raisebox{-0.3pt}[2pt]{\textbf{\orcidlink{0009-0002-3620-5481}}}}}},
S. Mohammad Razavizadeh\textsuperscript{\textsf{\scalebox{1.3}{\raisebox{-0.3pt}[2pt]{\textbf{\orcidlink{0000-0002-1732-4779}}}}}},~\IEEEmembership{Senior Member,~IEEE,} and
Halim Yanikomeroglu\textsuperscript{\textsf{\scalebox{1.3}{\raisebox{-0.3pt}[2pt]{\textbf{\orcidlink{0000-0003-4776-9354}}}}}},~\IEEEmembership{Fellow,~IEEE}}%
\thanks {Reza Ghasemi Alavicheh and S. M. Razavizadeh are with the School of Electrical Engineering, Iran University of Science and Technology (IUST), Tehran, Iran (emails: {\href{mailto:ghasemi_r@alumni.iust.ac.ir}{ghasemi{\textunderscore}r@alumni.iust.ac.ir} and \href{mailto:smrazavi@iust.ac.ir}{smrazavi@iust.ac.ir}}).

H. Yanikomeroglu is with the Non-Terrestrial Networks (NTN) Lab, Department of System and Computer Engineering (SCE), Carleton University, Ottawa, Canada (email: \href{mailto:halim@sce.carleton.ca}{halim@sce.carleton.ca}). \\ CORRESPONDING AUTHOR: S. Mohammad Razavizadeh }
}

\maketitle
\begin{abstract}
	In this paper, we explore the problem of utilizing Integrated Access and Backhaul (IAB) technology in Non-Terrestrial Networks (NTN), with a particular focus on aerial access networks. We consider an Uncrewed Aerial Vehicle (UAV)-based wireless network comprised of two layers of UAVs: (a) a lower layer consisting a number of flying users and a UAV Base Station (BS) that provides coverage for terrestrial users and, (b) an upper layer designated to provide both wireless access for flying users and backhaul connectivity for UAV BS. By adopting IAB technology, the backhaul and access links collaboratively share their resources, enabling aerial backhauling and the utilization of the same infrastructure and frequency resources for access links. A sum-rate maximization problem is formulated by considering aerial backhaul constraints to optimally allocate the frequency spectrum between aerial and terrestrial networks. We decompose the resulting non-convex optimization problem into two sub-problems of beamforming and spectrum allocation and then propose efficient solutions for each. Numerical results in different scenarios yield insightful findings about the effectiveness of using the IAB technique in aerial networks.
\end{abstract}
\section{Introduction} \label{sec:I}
We are witnessing a significant increase in the demands for wireless services and applications, leading to more densely populated networks of users and Base Stations (BS). For instance, BS density has increased from $5~\rm{BSs}/\rm{km}^2$ in 3G to over $50~\rm{BSs}/\rm{km}^2$ in 5G, and the deployment plans are shifting from Macro Base Stations (MBS) to Small Base Stations (SBS) to manage the massive traffic loads \textcolor{blue}{\cite{ge20165g}}. This growth will place immense stress on the Radio Access Network (RAN) and especially the backhaul links. Employing fiber optic backhaul for each SBS is both economically and geographically impractical. Therefore, new generations of wireless networks like 6G require cost-effective wireless backhaul solutions for practicality, scalability, and reduced hardware \textcolor{blue}{\cite{ni2019enhancing}}.

Integrated Access and Backhaul (IAB), first proposed in 3GPP Release 15 \textcolor{blue}{\cite{Specification2014}} is a promising wireless backhauling technique that addresses the challenges of dense mobile networks through self-backhauling, combining access and backhaul link resources in the BS. It uses a chain architecture, where multiple SBSs are ultimately connected to a macro BS.
Traditional wireless backhaul in microwave frequency bands is unsuccessful due to insufficient spectrum availability to meet the ever-evolving capacity demands. In contrast, millimeter-Wave (mmWave) bands offer a cost-effective solution for high-speed data transmission, making them ideal for wireless backhaul. However, the mmWave bands suffer from high path-loss and range restrictions.
The IAB technique addresses these challenges with the help of this multi-hop capability and shared utilization of both access and backhaul resources \textcolor{blue}{\cite{madapatha2020integrated}}.
Although the IAB technique is well-understood as a potential solution for the backhaul limitations in dense terrestrial networks \textcolor{blue}{\cite{8493520}}, \textcolor{blue}{\cite{8882288}}, and \textcolor{blue}{{\cite{10177930}}} its usage in Aerial Networks (ANs) is still under exploration and requires further research and evaluation to provide better coverage for ANs \textcolor{blue}{\cite{chen2022dedicating}}.
In general, when studying the IAB technique and UAV-based networks together, two approaches can be adopted.
In the first approach, UAVs are used to help and improve the performance of the IAB technique in providing the required backhaul for terrestrial networks (\textit{UAV-assisted IAB}) \textcolor{blue}{\cite{fouda2019interference}},  \textcolor{blue}{\cite{tafintsev2020reinforcement}},  \textcolor{blue}
{\cite{fouda2018uav}}, \textcolor{blue}{\cite{tafintsev2020aerial}}, \textcolor{blue}{\cite{tafintsev2023airborne}}, \textcolor{blue}{\cite{zhang2023packet}}, \textcolor{blue}{\cite{diamanti2021prospect}}.
The second approach that (to the best of our knowledge) has not been explored in the literature is utilizing the IAB technique to address the backhaul challenges to the UAV-based ANs (\textit{IAB-assisted UAV}).
Regarding the first approach, the problem of mitigating interference between access and backhaul links in the UAV-assisted IAB networks has been studied in some papers. Using terrestrial BSs for backhaul and exploiting UAVs' 3D mobility for interference management and power optimization was studied in \textcolor{blue}{\cite{fouda2019interference}}, \textcolor{blue}{\cite{fouda2018uav}}. Some of the papers investigate the optimal placement of UAV-IAB nodes and proposed solutions such as adaptive deployment \textcolor{blue}{ \cite{tafintsev2020aerial}} and joint path selection and resource allocation \textcolor{blue}{\cite{tafintsev2023airborne}}.
Moreover, some works compare the performance of UAV-assisted IAB networks with fully ground-based networks in terms of energy efficiency \textcolor{blue}{\cite{zhang2023packet}}, \textcolor{blue}{\cite{diamanti2021prospect}}. Another aspect considered is the uplink access of users and imperfect fronthaul of UAV-aided IAB in a cell-free network, which can be optimized by adjusting UAV power and 3D placement \textcolor{blue}{\cite{diaz2023cell}}.
However, most of these works fail to address the challenge of limited aerial access and backhaul coverage caused by down-tilt antennas of terrestrial BSs, as noted in \textcolor{blue}{\cite{liu2019multi}}.

As discussed above, in the realm of aerial wireless backhaul, the concept of IAB-assisted UAV networks remains relatively unexplored. In this paper, we aim to fill this gap by proposing a novel framework for IAB-assisted UAV networks. The primary objective is to alleviate the backhaul bottleneck commonly encountered in UAV networks, which also has the potential to worsen and become critical in future wireless networks that rely more on non-terrestrial networks \textcolor{blue}{\cite{7744808}}. This is achieved by leveraging aerial IAB-nodes and eliminating reliance on terrestrial BSs.
{  While our proposed framework does consider both aerial and terrestrial network coverage as a result of its innovative architecture, and addresses interference issues through the application of a hybrid beamforming method, the main objective is to evaluate our proposed network’s performance in terms of network sum-rate. This evaluation is carried out using analytical methods and examined with simulation results.}
Our considered Uncrewed Aerial Systems (UAS) or aerial wireless network model employs a multi-tier aerial architecture, comprising two tiers of UAV BSs. In the lower tier, there is a UAV BS that provides coverage for terrestrial users. This UAV acts as a so-called \textit{IAB-node} for the IAB technique and receives aerial backhaul connectivity service from a so-called \textit{IAB-donor} in the upper layer. It is designated to provide aerial access for some flying users or aerial User Equipment (UE) as well as backhaul connectivity for UAV BS.
In addition, in the upper layer, the IAB-donor UAV node is assumed to be potentially part of a Vertical Heterogeneous Network (VHetNet) and connected to High Altitude Platform Stations (HAPS) or satellites for its backhaul provisioning. These aerial and terrestrial networks share the same frequency spectrum resources, and we aim to optimize the allocation of the spectrum between them, with the goal of achieving sum-rate maximization. To address the non-convex and complex optimization problem, we propose an efficient solution that decomposes the main problem into two sub-problems. We spatially separate the aerial access and backhaul links by proposing a hybrid beamforming scheme within this Multiple-Input Single-Output (MISO) multi-user system. Subsequently, we derive a closed-form solution for the spectrum allocation problem.
In conclusion, we compare the results obtained from our approach with an equivalent non-IAB scenario. Our findings underscore the superior performance of the IAB approach for aerial wireless backhauling, particularly in terms of sum-rate and energy efficiency, making it suitable for catering to both aerial and terrestrial users.

The paper is structured in the following way: We begin by presenting the system and channel models in Sec.~\ref{sec:II}. Then, in Sec.~\ref{sec:III}, we discuss the problem formulation and present the optimization problems, which include hybrid beamforming design, spectrum allocation, and evaluation of bandwidth impact on thermal noise. Finally, we present the numerical results in Sec.~\ref{sec:IV}, followed by the conclusion in Sec.~\ref{sec:V}.\\
\textit{\textbf{Notation Conventions}}: Symbol ${\mathbb{C}^{m \times n}} $ denotes a complex space with dimensions $\mathrm{m}$ by $\mathrm{n}$. Scalar values, vectors, and matrices are respectively represented by $\mathrm{a}$, $\textbf{a}$, and $\textbf{A}$. Transpose and Hermitian transpose of a matrix are indicated by $\textbf{A}^{T}$ and $\textbf{A}^{H}$. Symbol $\textbf{I}_K$ signifies an identity matrix of size $K \times K$. $\textbf{[A]}_{i,j}$ denotes the element in the $i$th row and $j$th column of matrix $\textbf{A}$. The symbols $\mathrm{Tr}\hspace{1pt}[.]$, $\mathrm{log(.)}$, and $\mathbb{E}\hspace{1pt}[.]$ represents the trace, logarithm, and expected value operations, respectively. A diagonal matrix is represented by $\mathrm{diag}\hspace{1pt}[.]$ Symbols ${\lvert . \rvert}$ and ${\lVert . \rVert}_F$ respectively denote absolute value and the Frobenius norm.

\subsection{IAB Workflow}
{   
Each node in IAB architecture consists of two components. In an IAB-node, these components are the IAB-node Distributed Unit (IAB-node-DU) and the IAB-node Mobile Termination (IAB-node-MT). In an IAB-donor, the components are the IAB-donor Centralized Unit (IAB-donor-CU) and IAB-donor-DU. The MT part is responsible for establishing the radio connection, including various backhaul link layers with the IAB-donor. The CU primarily handles the control plane and the user plane functionalities for the backhaul. The DU is responsible for providing New Radio (NR) functions to a UE or the MT part of another IAB-node. To better integrity with the traditional network model, based on \textcolor{blue}{\cite{TS2024}}, the MT part of each node acts like a UE, and the DU part functions like the DU of a gNB. A new protocol in IAB, called Backhaul Adaptation Protocol (BAP), is responsible for forwarding Internet protocol (IP)-based packets between an IAB-node or IAB-donor over the backhaul link. The BAP header includes a unique identifier that specifies the connection's destination IAB-node and the path (if there are multiple nodes ahead) through the BAP address and BAP path ID. The BAP header must ensure that the sent packets reach their final destination, distinguishing IAB backhaul features from the traditional model. The lower layer protocols, including physical (PHY), Medium Access Control (MAC), and Radio Link Control (RLC), and the upper layer protocols, including Packet Data Convergence Protocol (PDCP), Radio Resource Control (RRC), and Service Data Adaptation Protocol (SDAP), are performed in the DU and CU, respectively. The process of adding an IAB-node to the network is designed by 3GPP to be similar to the traditional method of adding a gNB to the network, which is carried out in three stages \textcolor{blue}{\cite{9963607}}, \textcolor{blue}{\cite{TS2021}}.

\vspace{0.6em}

\noindent{\textbf{Initialization and Synchronization:}} The IAB-node-MT initiates a connection to the IAB-donor similar to a regular UE, acting in the lower layer of the system model. This initialization and synchronization process involves the IAB-node-MT scanning the frequency band for Synchronization Signal Blocks (SSBs) and performing a random access procedure. Following this, an RRC connection is established with the IAB-donor-CU and authentication with the Core Network is completed.

\vspace{0.6em}

\noindent{\textbf{Backhaul Link Establishment:}} Next, the backhaul link is established. This involves configuring a new Backhaul RLC channel. The BAP address for the IAB-node is configured, and the default BAP Routing ID for the upstream direction is set. Additionally, the BAP sublayer is updated to support routing between the new IAB-node and the IAB-donor-DU, ensuring efficient data flow.

\vspace{0.6em}

\noindent{\textbf{IAB-node Configuration:}} Finally, the DU part of the new IAB-node is configured, and an F1 link is established between the IAB-donor-CU and the IAB-node-DU. Once setup is complete, the new IAB-node begins serving UEs.
}
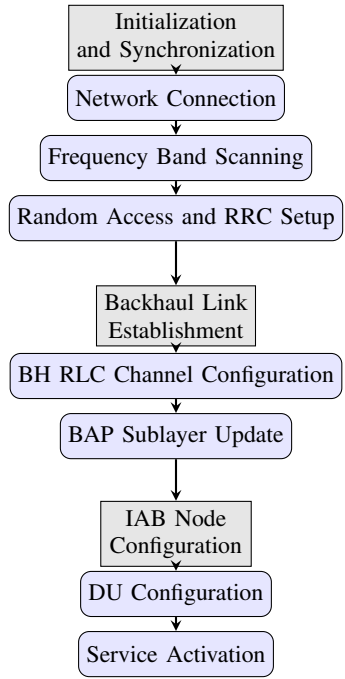
\begin{figure}[h!]
    \centering
    \begin{tikzpicture}[node distance=0.8cm, align=center]
        \footnotesize  
        
        \node (section1) [section] {Initialization\\and Synchronization};
        \node (net) [process, below of=section1] {Network Connection};
        \node (scan) [process, below of=net] {Frequency Band Scanning};
        \node (rrc) [process, below of=scan] {Random Access and RRC Setup};
        
        \node (section2) [section, below of=rrc, yshift=-0.5cm] {Backhaul Link\\Establishment};
        \node (bhRLC) [process, below of=section2] {BH RLC Channel Configuration};
        \node (bap) [process, below of=bhRLC] {BAP Sublayer Update};
        
        \node (section3) [section, below of=bap, yshift=-0.5cm] {IAB Node\\Configuration};
        \node (du) [process, below of=section3] {DU Configuration};
        \node (service) [process, below of=du] {Service Activation};
        
        \draw [arrow] (section1) -- (net);
        \draw [arrow] (net) -- (scan);
        \draw [arrow] (scan) -- (rrc);
        
        \draw [arrow] (rrc) -- (section2);
        \draw [arrow] (section2) -- (bhRLC);
        \draw [arrow] (bhRLC) -- (bap);
        
        \draw [arrow] (bap) -- (section3);
        \draw [arrow] (section3) -- (du);
        \draw [arrow] (du) -- (service);
        
    \end{tikzpicture}
    \caption{Three stages of IAB Workflow.}
    \label{fig:workflow}
\end{figure}
\section{System Model}\label{sec:II}
\begin{figure}[tp]
	\centering
	\includegraphics[width=0.5\textwidth]{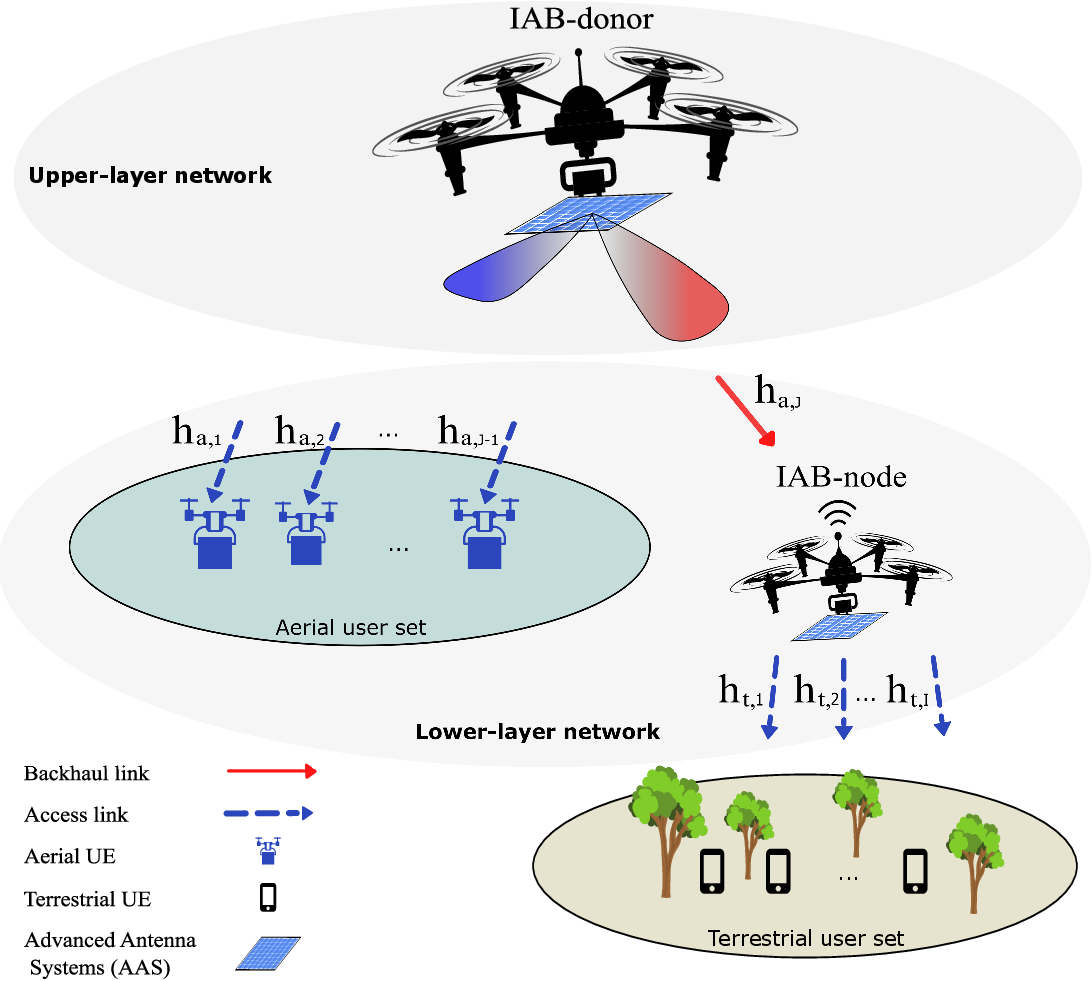}
	\caption{Aerial Integrated Access and Backhaul (IAB) enabled system model serving two sets of aerial and terrestrial users.}
	\label{fig:system_model}
\end{figure}

\begin{figure}[tp]
	\centering
	\includegraphics[width=0.5\textwidth]{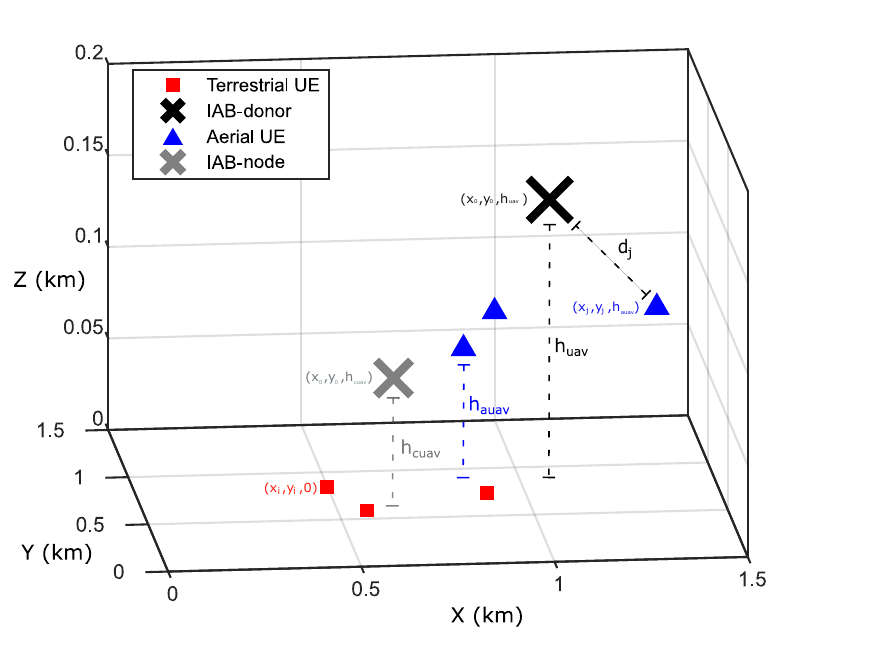}
	\caption{In a $2.25~\textrm{km}^2$ area, both aerial and terrestrial UEs are randomly distributed according to a normal distribution. They satisfy a minimum specified distance from each other, as detailed in \textcolor{blue}{\cite{shehzad2021backhaul}}, and are within defined sub-regions. All aerial UEs are located at the same altitude.}
	\label{fig:system_model_simulation}
\end{figure}

The proposed model can be incorporated into a broader Non-Terrestrial Network (NTN) scenario, which includes satellites, HAPS, and additional UAVs, for instances such as urban emergency coverage \textcolor{blue}{\cite{9380673}} and { situations like natural disasters or other emergency events in urban areas where terrestrial communication infrastructure may be compromised. The proposed system model leverages an upper-layer of aerial nodes (e.g. UAV) to provide backhaul connectivity to a lower-layer of aerial base stations serving both aerial and terrestrial users. During natural disasters or emergencies when terrestrial networks may fail, maintaining this aerial backhaul connectivity becomes critical to ensure data can still be forwarded between the upper aerial layer and terrestrial users served by the lower-layer IAB-node. Data for terrestrial users must be forwarded through the aerial backhaul link from the upper-layer network (IAB-donor) to the lower-layer aerial base station (IAB-node), as the IAB-node doesn't have a direct connection to the core network. The flexibility of the IAB approach to dynamically allocate bandwidth and resources between aerial and terrestrial links (as opposed to static allocation in non-IAB) can help alleviate the backhaul bottleneck when more capacity is needed.}
Each IAB-node is equipped with a uniform planar array (UPA) with total $N=N_x\times N_y$ array elements. Each antenna array is divided into multiple sub-arrays, in which some are assigned to the aerial access links and others to the aerial backhaul link in the IAB-donor node. The LoS aerial channel between an aerial user and the IAB-donor and also the LoS backhaul channel are denoted by ${\mathbf{h}}_{a,j} \in \mathbb{C}^{N \times 1} $. Also, the channel between a terrestrial user and the IAB-node is represented by ${\mathbf{h}}_{t,i} \in \mathbb{C}^{N \times 1} $. IAB-donor and IAB-node are associated with a fixed number of users from $J-1$ and $I$, respectively, as shown in Fig.~\ref{fig:system_model}. The network comprises a total of $J$ aerial-to-aerial links, with $J-1$ links dedicated to aerial users and one link serving as the backhaul for the IAB-node. Additionally, there are $I$ aerial-to-terrestrial links.
Since we were considering a mmWave scenario, we used the Saleh-Valenzuela channel model \textcolor{blue}{\cite{feng2021hybrid}}, \textcolor{blue}{\cite{varshney2022design}}, \textcolor{blue}{\cite{6717211}} that can be expressed as follows for aerial and terrestrial connections:
\begin{table}[tp]
\caption{The Definitions of The Symbols}
\centering
\resizebox{\columnwidth}{!}{%
\begin{tabular}{|c|c|}
\hline
\textbf{Symbol} & \textbf{Definition} \\ \hline
$N$ & Total number of array elements in the uniform planar array (UPA) \\ \hline
$N_x \times N_y$ & Dimensions of the UPA \\ \hline
${\mathbf{h}}_{a,j}$ & Channel between an aerial user or IAB-node and the IAB-donor \\ \hline
${\mathbf{h}}_{t,i}$ & Channel between a terrestrial user and the IAB-node \\ \hline
$J$ & Number of aerial-to-aerial links \\ \hline
$I$ & Number of aerial-to-terrestrial links \\ \hline
$\gamma_0$ & Channel power gain or reference path loss \\ \hline
$d_0$ & Reference distance \\ \hline
$n_{LoS}$ & Path loss exponent \\ \hline
$d_j$ & Euclidean distance between an aerial user and the IAB-donor \\ \hline
$d_i$ & Euclidean distance between a terrestrial user and the IAB-node \\ \hline
$\theta, \phi$ & Azimuth and elevation angles representing the array's Angle of Departure (AoD) \\ \hline
$\mathbf{a}(\theta, \phi)$ & Array steering vector \\ \hline
$r$ & Antenna element spacing, expressed in half-wavelength \\ \hline
$\lambda$ & Wavelength \\ \hline
$\mathbf{F}_a^{RF}, \mathbf{F}_t^{RF}$ & Analog beamformers for aerial and terrestrial links \\ \hline
$\mathbf{F}_a^{BB}, \mathbf{F}_t^{BB}$ & Digital beamformers for aerial and terrestrial links \\ \hline
$\Lambda$ & Frobenius norm between $\mathbf{F}^{zf}$ and $\mathbf{F}^{RF}\mathbf{F}^{BB}$ \\ \hline
$N_{RF}$ & Number of Radio Frequency (RF) chains \\ \hline
$p_j, p_i$ & Transmission power for aerial and terrestrial users \\ \hline
$s_{a,j}, s_{t,i}$ & Transmitted symbols for aerial and terrestrial users \\ \hline
$P_1, P_2$ & Total available power at IAB-donor and IAB-node \\ \hline
$n_a, n_t$ & Thermal noise for aerial and terrestrial links \\ \hline
$P_n$ & Thermal noise power spectral density \\ \hline
$\mathbf{P}_t, \mathbf{P}_a$ & Diagonal power allocation matrix for terrestrial and aerial users \\ \hline
$\mathbf{I}_a, \mathbf{I}_t$ & Intra-tier interference terms for aerial and terrestrial links \\ \hline
$w$ & Total available bandwidth \\ \hline
$\mu_a, \mu_t$ & Bandwidth coefficients for aerial and terrestrial links \\ \hline
$R_{a,j}, R_{t,i}$ & Received Spectral Efficiency (SE) for aerial and terrestrial links \\ \hline
$\mathfrak{R}_a, \mathfrak{R}_t$ & Sum-SE of aerial and terrestrial users \\ \hline
$\sigma_a^2, \sigma_t^2$ & Thermal noise power for aerial and terrestrial links \\ \hline
$\mathfrak{R} $ & Total sum-rate of the network \\ \hline
$\zeta$ & Bottleneck rate of aerial backhaul link \\ \hline
\end{tabular}}
\label{tab:variables}
\end{table}
\begin{subequations}
	\begin{align}
		\label{eq:steeringvectorcross1}
		\mathbf{h}_{{a,j}}    & =\sqrt{\frac{\gamma_{0}}{{d_{j}}^{n_{LoS}}}}
		\mathbf{a}\left(\theta_j^{{a}}, \phi_j^{{a}}\right),               \\
		\label{eq:steeringvectorcross2}
		\mathbf{h}_{{t}, {i}} & =\sqrt{\frac{\gamma_{0}}{{d_{i}}^{n_{LoS}}}}
		\mathbf{a}\left(\theta_i^{{t}}, \phi_i^{{t}}\right).
	\end{align}
\end{subequations}
The channel matrix, containing all the channel vectors of each set of users, can be expressed as:
\begin{subequations}
	\begin{align}
		\mathbf{H}_{{a}} & =[\mathbf{h}_{{a,1}} \space \vdots \space \dots \space \vdots
		\space  \mathbf{h}_{a,J}]^T, \hspace{0.5 cm}   \mathbf{H}_{a} \in \mathbb{C}^{J \times N },
		\\
		\mathbf{H}_{{t}} & = [\mathbf{h}_{{t,1}} \space \vdots \space \dots \space \vdots
		\space  \mathbf{h}_{t,I}]^T, \hspace{0.65 cm}   \mathbf{H}_{t} \in \mathbb{C}^{I \times N }.
	\end{align}
\end{subequations}
According to \textcolor{blue}{(\ref{eq:steeringvectorcross1})} and \textcolor{blue}{(\ref{eq:steeringvectorcross2})}, $\gamma_{0} = -20\log_{10}(f) - 32.45$ [dB] represents the channel power gain or reference path loss with $d_{0} = 1$ m as a reference distance \textcolor{blue}{\cite{jeonEnergyEfficientAerialBackhaul2022b}}. The constant $n$ denotes the path loss exponent and its value typically falls within the range of $2-4$. {  Fig.~\ref{fig:system_model_simulation} illustrates a 3D simulation environment representing the proposed system model. The positions of the nodes are denoted by their Cartesian coordinates $(x, y, z)$, while their heights are represented by $h$. The Euclidean distances separating the nodes are represented by $d$. Specifically, the heights of the IAB-donor and IAB-node are denoted as $h_{\text{uav}}$ and $h_{\text{cuav}}$, respectively. The aerial users are positioned at a height of $h_{j} = h_{\text{auav}}$, whereas the terrestrial users are considered to be at ground level, with a height of $h_i = 0$.
} The distances $d_{j}$ and $d_{i}$ represent the separation between an aerial user and the IAB-donor or between a terrestrial user and the IAB-node, as described below:
\begin{subequations}
\begin{align}
	\label{eq:distance}
	{d}_{j} &= \sqrt{({x}_j - {x}_0)^2 + ({y}_j - {y}_0)^2 + ({h}_j - {h}_{uav} )^2 }, \\
        \label{eq:distance2}
	{d}_{i} &= \sqrt{({x}_i - {x}_0)^2 + ({y}_i - {y}_0)^2 + ({h}_i - {h}_{cuav} )^2 }.
\end{align}
\end{subequations}
In the azimuth and elevation directions, $\phi$ and $\theta$ respectively represent the array's Angle of Departure (AoD). The array steering vector for both the aerial and terrestrial networks can be expressed as
\begin{align}
	\label{eq:steeringvectorcross_final}
	\mathbf{a}\left(\theta, \phi\right) & = [1, \space e^{j\frac{2\pi}{\lambda}r\sin{(\theta)}[(n_x-1)\cos{(\phi)}+(n_y-1)\sin{(\phi)}]} \space,
	\cdots, \notag                                                                                                                               \\
	                                    & \hspace{1cm} e^{j\frac{2\pi}{\lambda}r\sin{(\theta)}[(N_x-1)\cos{(\phi)}+(N_y-1)\sin{(\phi)}]} ]^T.
\end{align}
The parameter $r$ in \textcolor{blue}{(\ref{eq:steeringvectorcross_final})} represents the antenna element spacing, expressed in half-wavelength. The downlink transmission of a hybrid analog-digital beamforming scheme is considered, where the analog and digital beamformers are denoted ${\mathbf{F}_a}^{RF}$, ${\mathbf{F}_t}^{RF} \in
	\mathbb{C}^{N \times N_{RF}} $ for both aerial and terrestrial links and ${\mathbf{F}}_a^{BB}
	\in \mathbb{C}^{N_{RF} \times J} $ for aerial access and backhaul links and ${\mathbf{F}}_t^{BB}
	\in \mathbb{C}^{N_{RF} \times I} $ for terrestrial access links in a Multiple Input Single Output (MISO) scenario \textcolor{blue}{\cite{majidzadeh2017partially}}, \textcolor{blue}{\cite{sohrabi2016hybrid}}.
$N_{RF}$ is the number of Radio Frequency (RF) chains that we consider to be equal to the number of users in each group. Digital or Base Band (BB) beamformers can be represented as follows:
\begin{subequations}
	\begin{align}
		\label{eq:BB_BF}
		\mathbf{F}_a^{BB} & = [\mathbf{f}_1^{BB} \ \dots \mathbf{f}_J^{BB}], \\
		\mathbf{F}_t^{BB} & = [\mathbf{f}_1^{BB}  \dots  \mathbf{f}_I^{BB}].
	\end{align}
\end{subequations}
By considering the Frequency Division Multiplexing (FDM) configuration, frequency resources are allocated to aerial and terrestrial access links. This allotment results in only intra-tier interference occurring solely between the aerial access and backhaul links, as well as the terrestrial access links. Also, hybrid analog-digital beamforming is employed to shape the radiation pattern of subarrays, direct energy to the desired direction, and minimize interference between links. This approach facilitates the establishment of multiple connections simultaneously. The received signals at the $j^{th}$ aerial (access or backhaul) link can be expressed as:
\begin{align}
	\label{eq:sig_aerial}
	y_{a, j}  = \sqrt{p_j} & \hspace{1 mm} \mathbf{h}_{a, j}^H \mathbf{F}_{a}^{R F} \mathbf{f}_{a, j}^{B B} s_{a, j} \\ &+
	\sum_{k=1/ j}^J \sqrt{p_k} \hspace{1 mm} \mathbf{h}_{a, j}^H \mathbf{F}_{a }^{R F} \mathbf{f}_{a, k}^{B B} s_{a, k} + n_{a}, \notag
\end{align}
and the received access signal at $i^{th}$ terrestrial user as:
\begin{align}
	\label{eq:sig_terrestrial}
	y_{t, i}  = \sqrt{p_i} & \hspace{1 mm} \mathbf{h}_{t, i}^H \mathbf{F}_{t}^{R F} \mathbf{f}_{t, i}^{B B} s_{t, i} \\ &+
	\sum_{k=1 / i}^I \sqrt{p_k} \hspace{1 mm} \mathbf{h}_{t, i}^H \mathbf{F}_{t}^{R F} \mathbf{f}_{t, k}^{B B} s_{t, k}+n_{t}. \notag
\end{align}
In these equations, the second term represents the interference among access links of users and the third term accounts for temperature noise. The transmitting power for each set is represented by a diagonal matrix for terrestrial links  $\mathbf{P}_{t}= $ \textrm{diag}($p_1$, ... ,$p_I$) and for aerial links $\mathbf{P}_{a}= $ \textrm{diag}($p_1$, ... ,$p_J$) consisting of power assigned to each link as $p_i = \frac{P_2}{I} $, $p_j = \frac{P'_1}{J-1} $ and $P'_1$ is the remaining power of IAB-donor after allocating the required power to the backhaul when IAB-node is detected among aerial users. Eq.~\textcolor{blue}{(\ref{eq:sig_aerial})} and \textcolor{blue}{(\ref{eq:sig_terrestrial})}
can be represented as a matrix scheme containing total signals for each group of users. In the below equations, interference terms are shown by $\mathbf{I}_a, \mathbf{I}_t$ and include all intra-tier interference between aerial links and terrestrial links and the transmitted symbols are ${\mathbf{s}}_{a} \in \mathbb{C}^{J \times 1}$ and ${\mathbf{s}}_{t} \in \mathbb{C}^{I \times 1}$ and $\mathbb{E}[{|\mathbf{s}}_{a}|^2] = 1$, $\mathbb{E}[{|\mathbf{s}}_{t}|^2] = 1$,
\begin{align}
	\mathbf{y}_{{a}} & =\mathbf{H}_{{a}} \mathbf{F}_{{a}}  \mathbf{s}_{{a}}+\mathbf{I}_{{a}}+\mathbf{n}_{{a}}, \\
	\mathbf{y}_{{t}} & =\mathbf{H}_{{t}} \mathbf{F}_{{t}} \mathbf{s}_{{t}}+\mathbf{I}_{{t}}+\mathbf{n}_{{t}}.
\end{align}
There is a total available bandwidth of $w$, and we aim to utilize this bandwidth efficiently to provide both aerial and terrestrial services. The bandwidth allocated to aerial (access and backhaul) links and terrestrial access links represented as $\mu_{a}$ and $\mu_{t}$, referred to as bandwidth coefficients, and they must satisfy the following conditions:
\begin{subequations}
\begin{align}
     &0 \leq \mu_a \leq 1, \\
     &0 \leq \mu_t \leq 1, \\
     &\mu_t + \mu_a = 1.
\end{align}
\end{subequations}
Received Spectral Efficiency (SE) at aerial and terrestrial links can be written as \textcolor{blue}{(\ref{eq:thr_a})} and \textcolor{blue}{(\ref{eq:thr_t})}
\begin{subequations}
	\begin{align}
		\label{eq:thr_a}
		R_{a, j} & =\log_{2} \left(1+\frac{{p_j}\left|\mathbf{h}_{a, j}^H \mathbf{F}_{a}^{R F}
			\mathbf{f}_{a, j}^{B B}\right|^2}{\sum_{k=1 / j}^J {p_k} \left|\mathbf{h}_{a, j}^H \mathbf{F}_{a}^{R F}
		\mathbf{f}_{a, k}^{B B}\right|^2+\sigma_{a}^2}\right),                                          \\
		\label{eq:thr_t}
		R_{t, i} & =  \log_{2} \left(1+\frac{{p_i} \left|\mathbf{h}_{t, i}^H \mathbf{F}_{t}^{R F}
			\mathbf{f}_{t, i}^{B B}\right|^2}{\sum_{k=1 / i}^I {p_k} \left|\mathbf{h}_{t, i}^H \mathbf{F}_{t}^{R F}
			\mathbf{f}_{t, k}^{B B}\right|^2+\sigma_{t}^2}\right).
	\end{align}
\end{subequations}
{  In these equations, $\sigma_a^2$ and $\sigma_t^2$ represent the thermal noise power for aerial and terrestrial links, respectively.}

\section{Problem Formulation}\label{sec:III}
Our objective is to maximize the total sum-rate of the network, which entails maximizing the sum-rate for both the aerial and terrestrial groups of users. The sum-rate can be expressed as:
\begin{subequations}
	\begin{align}
		\label{eq:rate_a_b}
		\mathfrak{R} & =\left(\mu_a\right) w\sum_{j=1}^{J-1}
		R_{a, j} +  \zeta,                                         \\
		\label{eq:rate_b_t}
		\zeta        & = \min \left\{\left(\mu_a\right) w R_{a,b}
		, \left(\mu_t\right) w\sum_{i=1}^I R_{t, i}\right\}.
	\end{align}
\end{subequations}
Eq.~\textcolor{blue}{(\ref{eq:rate_a_b})} consists of two terms: The first term represents the sum-rate of aerial users, while the second term represents the sum-rate of terrestrial users. In \textcolor{blue}{(\ref{eq:rate_b_t})}, aerial backhaul rate denoted as $R_{a,b}$, acts as a bottleneck and limits the rate of terrestrial users. {  This effect stems from the fact that in the proposed architecture, the data to terrestrial users gets forwarded through the aerial backhaul link from the upper-layer network (IAB-donor) to the lower-layer aerial base station (IAB-node). So the capacity of this aerial backhaul link can limit the overall communication rate for terrestrial users.} {{
As the backhaul constraint is related to the value of the bandwidth coefficient, it is important to take into account its effects on the bandwidth coefficient range. We initially accounted for the backhaul limitation through the minimum operation in the objective function \textcolor{blue}{(\ref{eq:rate_b_t})}. However, to make the backhaul constraint and its effect on bandwidth coefficients more explicit, it is introduced as a separate constraint further in equation \textcolor{blue}{(\ref{eq:c_backhaul__after_bf})}, and these constraints are investigated in Sec.~\textcolor{blue}{}\ref{sec:spec}.}}
With the objective function $\mathfrak{R}$, the problem of maximizing the sum-rate, considering bandwidth allocation, hybrid beamforming, and power limitations of flying BSs, can be formulated as follows in \textcolor{blue}{(\ref{eq:objective})}. $\mathbf{F}_{a}$ and $\mathbf{F}_{t}$ are hybrid beamforming matrices, constraints \textcolor{blue}{(\ref{eq:c_power_a})} and \textcolor{blue}{(\ref{eq:c_power_t})} define the total power limitation and $\mathrm{P_1}$ and $\mathrm{P_2}$ indicate the available power at each UAV BS (IAB-donor and IAB-node).
Given that phase shifters in analog beamformers only change the phase of signals, each element of the analog matrix should have a constant absolute value in constraints \textcolor{blue}{(\ref{eq:c_RF_a})} and \textcolor{blue}{(\ref{eq:c_RF_t})}. The height and intra-distance of UAV-IABs are expressed as $\mathrm{h_{uav}}$ and $\mathrm{d_{uav}}$ and are considered to be unchanging and assuming the same priority for all users.
Due to the complexity and non-convexity of the optimization problem, caused by \textcolor{blue}{(\ref{eq:c_RF_a})} and \textcolor{blue}{(\ref{eq:c_RF_t})}, we aim to separate the design of hybrid beamforming from the optimization of bandwidth coefficients, this decoupling is achieved by assuming fixed bandwidth coefficients, allowing us to independently maximize the SE in the first and second terms of the objective function
\begin{subequations}
	\begin{align}
		\max_{\substack{\mu_a, \mu_t, \mathbf{F}_{a}^{RF},\mathbf{F}_{t}^{RF},                                                                                       \\ \mathbf{F}_{a}^{BB}, \mathbf{F}_{t}^{BB}}}
		            & \label{eq:objective} \hspace{1mm} \mathfrak{R}(\mu_a, \mu_t, \mathbf{F}_{a}^{RF},\mathbf{F}_{t}^{RF},\mathbf{F}_{a}^{BB}, \mathbf{F}_{t}^{BB}) \\
		\text{s.t.} & \label{eq:ca} \quad 0 \leq \mu_a \leq 1,                                                                                                           \\
		            & \label{eq:ct} \quad 0 \leq \mu_t \leq 1,
		\\
		            & \label{eq:sum_mu} \quad  \mu_t + \mu_a = 1,                                                                                                    \\
		            & \label{eq:c_RF_a} \quad \left|\left[\mathbf{F}_{a}^{R F}\right]_{l, l^{\prime}}\right|=1 \hspace{6pt} \forall l, l^{\prime},                         \\
		            & \label{eq:c_RF_t} \quad \left|\left[\mathbf{F}_{t}^{R F}\right]_{l, l^{\prime}}\right|=1 \hspace{6pt} \forall l, l^{\prime},                          \\
		            & \label{eq:c_power_a} \quad   \operatorname{Tr}(\mathbf{{F_a}^{H}{F_a}})\le \mathrm{P_1},                                                        \\
		            & \label{eq:c_power_t} \quad   \operatorname{Tr}(\mathbf{{F_t}^{H}{F_t}})\le \mathrm{P_2}.
	\end{align}
\end{subequations}
The problem for hybrid beamforming design can be rewritten as follows:
\begin{subequations}
	\begin{align}
		\max_{\substack{\mathbf{F}_{a}^{RF},\space  	\mathbf{F}_{a}^{BB}}}
		            & \label{eq:objective_2}  \sum_{j=1}^J  \log_{2} \left(1+\frac{{p_j} \left|\mathbf{h}_{a, j}^H \mathbf{F}_{a}^{R F} \mathbf{f}_{a, j}^{B B}\right|^2}{\sum_{k=1 / j}^J {p_k} \left|\mathbf{h}_{a, k}^H \mathbf{F}_{a}^{RF} \mathbf{f}_{a, k}^{BB}\right|^2+\sigma_{a}^2}\right) \\
		\text{s.t.} & \label{eq:c_RF_a_2} \quad \left|\left[\mathbf{F}_{a}^{R F}\right]_{l, l^{\prime}}\right|=1 \hspace{5mm}\forall l, l^{\prime},                                                                                                                                                            \\
		            & \label{eq:c_power_2} \quad \operatorname{Tr}(\mathbf{F}_a^H\mathbf{F}_a)\le \mathrm{P_1},
	\end{align}
\end{subequations}
\begin{subequations}
	\begin{align}
		\max_{\substack{\mathbf{F}_{t}^{RF},\space \mathbf{F}_{t}^{BB}}}
		            & \label{eq:objective_3}  \sum_{i=1}^I  \log_{2} \left(1+\frac{{p_i} \left|\mathbf{h}_{t, i}^H \mathbf{F}_{t}^{R F} \mathbf{f}_{t, i}^{BB}\right|^2}{\sum_{k=1 / i}^I {p_k} \left|\mathbf{h}_{t, k}^H \mathbf{F}_{t}^{RF} \mathbf{f}_{t, k}^{BB}\right|^2+\sigma_{t}^2}\right) \\
		\text{s.t.} & \label{eq:c_RF_t_3} \quad \left|\left[\mathbf{F}_{t}^{R F}\right]_{l, l^{\prime}}\right|=1 \hspace{5mm}\forall l, l^{\prime},                                                                                                                                                           \\
		            & \label{eq:c_power_3} \quad \operatorname{Tr}(\mathbf{F}_t^H\mathbf{F}_t)\le \mathrm{P_2}.
	\end{align}
\end{subequations}
According to \textcolor{blue}{(\ref{eq:objective_2})} and \textcolor{blue}{(\ref{eq:objective_3})}, two non-convex optimization problems are challenging to solve.
As an alternative approach, we can assume that the zero-forcing fully digital beamformer as described in \textcolor{blue}{(\ref{eq:zf})} is the upper bound performance for our problems, and we can design our hybrid beamforming compared to zero-forcing beamforming. Before solving alternative problems, the total available power $P_1$ and $P_2$ was equally shared among users. In both aerial and terrestrial user sets, problems \textcolor{blue}{(\ref{eq:objective_2})} and \textcolor{blue}{(\ref{eq:objective_3})} can be expressed as follows:
\begin{subequations}
	\begin{align}
		\max_{\substack{\mathbf{F}^{RF}, \mathbf{F}^{BB}}}
		            & \label{eq:objective_4} \quad   R(\mathbf{F}^{RF},\mathbf{F}^{BB})                                                      \\
		\text{s.t.} & \label{eq:c_RF_4} \quad \left|\left[\mathbf{F}^{R F}\right]_{l, l^{\prime}}\right|=1 \hspace{5mm}\forall l, l^{\prime}, \\
		            & \label{eq:c_power_4} \quad \operatorname{Tr}(\mathbf{F}^H\mathbf{F})\le \mathrm{P_d}.
	\end{align}
\end{subequations}
\subsection{Problem of Hybrid Precoders}
An upper bound performance can be obtained using a zero-forcing beamforming technique \textcolor{blue}{\cite{majidzadeh2017partially}}, as shown in the following:
\begin{align}
	\label{eq:zf}
	\mathbf{F}^{zf} = \mathbf{H}^H (\mathbf{H}\mathbf{H}^H)^{-1}.
\end{align}
Eq. \textcolor{blue}{(\ref{eq:objective_4})} can be adapted to use either terrestrial or aerial channels, along with their respective power allocation matrices. We aim to design hybrid digital and analog precoders that are as close as possible to the zero-forcing precoder, denoted as $\Lambda = {\lVert \mathbf{F}^{zf} - \mathbf{F}^{RF} \mathbf{F}^{BB}\rVert}_F $. Therefore, it is possible to formulate this problem as follows:
\begin{subequations}
	\begin{align}
		\min_{\substack{\mathbf{F}^{RF}, \mathbf{F}^{BB}}}
		            & \label{eq:objective_fro} \quad    {\lVert \mathbf{F}^{zf} - \mathbf{F}^{RF} \mathbf{F}^{BB}\rVert}_F                     \\
		\text{s.t.} & \label{eq:c_RF_fro} \quad \left|\left[\mathbf{F}^{R F}\right]_{l, l^{\prime}}\right|=1 \hspace{5mm}\forall l, l^{\prime}.
	\end{align}
\end{subequations}
The elements of the analog beamformer matrix are generated using an exponential function $[\mathbf{F}^{R F}]_{l, l^{\prime}} = e^{j{\theta}_{l,l{\prime}}},  \hspace{2 mm} \forall l, l^{\prime}$ with unit absolute values, $l$ and $l^{\prime}$ represent the row and column of the analog beamforming matrix; we optimize the digital beamforming matrix and the phases of the analog beamformer matrix in two separate convex optimization problems by keeping one of the optimization variables fixed and iteratively optimizing the other until the objective function meets a predefined criterion. With the use of a fixed digital beamformer, we can express the problem \textcolor{blue}{(\ref{eq:objective_4})} in the following manner:
\begin{align}
	\min_{\substack{\mathbf{F}^{RF}}}
	 & \label{eq:part_2_opt_fro} \quad    {\lVert \mathbf{F}^{zf} - \mathbf{F}^{RF} \mathbf{F}^{BB}\rVert}_F.
\end{align}
Initially, the unit value constraint of the RF precoder is temporarily ignored, and the RF precoder matrix is computed without this constraint. Subsequently, its elements are normalized as expressed in \textcolor{blue}{(\ref{eq:Update_RF})}, then fix the analog beamformer turns optimization \textcolor{blue}{(\ref{eq:objective_4})} into an unconstrained and convex problem, as shown below:
\begin{align}
	\min_{\substack{\mathbf{F}^{BB}}}
	 & \label{eq:part_1_opt_fro} \quad    {\lVert \mathbf{F}^{zf} - \mathbf{F}^{RF} \mathbf{F}^{BB}\rVert}_F,
\end{align}
\RestyleAlgo{ruled}
\SetKwComment{Comment}{\%}{}
\begin{algorithm}[t]
	\SetKwInOut{Input}{Input}
	\SetKwInOut{Output}{Output}
	\SetKwInOut{Return}{Return}
	\caption{Hybrid Beamforming for Aerial and Terrestrial Networks}\label{al:1}
	 \Input{Initial $\mathbf{F}^{BB}$ and  $\mathbf{F}^{RF}$ by SVD of $\mathbf{F}^{ZF}$ in \textcolor{blue}{(\ref{eq:initial})} and \textcolor{blue}{(\ref{eq:initial2})} and update $\mathbf{F}^{RF}$ using \textcolor{blue}{(\ref{eq:Update_RF})}. {$\ell = J$ for aerial and $\ell = I$ for terrestrial set.}}
	\Output{ $\mathbf{F}^{BB},  \mathbf{F}^{RF}, \mathbf{F}_{hyb}$.}

	\For{$k = 1 ...  K$}{
		
			Fix $\mathbf{F}^{BB}$\;
			$\mathbf{F}^{RF}  \gets \rm{arg~min}{\lVert \mathbf{F}^{zf} - \mathbf{F}^{RF}\mathbf{F}^{BB}\rVert}_F$\;
			Update $\mathbf{F}^{RF} $ using \textcolor{blue}{(\ref{eq:Update_RF})}\;
			
			Fix $\mathbf{F}^{RF}$\;
			$\mathbf{F}^{BB}  \gets \rm{arg~min}{\lVert \mathbf{F}^{zf} - \mathbf{F}^{RF}\mathbf{F}^{BB}\rVert}_F$\;
			\eIf{${g}_{k} - {g}_{k+1} \leq \text{criterion}$ or max iter. happens}
				{$\mathbf{P} = \rm{diag}(p_{1}, ..., p_{\ell}$)\;

				$\mathbf{F}_{hyb} \gets \frac{\mathbf{F}_{RF} \mathbf{F}_{BB}}{{\lVert {\mathbf{F}_{RF} \mathbf{F}_{BB}\rVert}_F}}\mathbf{P}^{\frac{1}{2}} $\;
				break;}
			{continue;}
			
		}
\end{algorithm}
{  The transition from \textcolor{blue}{(\ref{eq:part_2_opt_fro})} to \textcolor{blue}{(\ref{eq:part_1_opt_fro})}, in conjunction with Algorithm~\textcolor{blue}{\ref{al:1}}, involves a two-stage iterative optimization for hybrid beamforming design.
In \textcolor{blue}{(\ref{eq:part_2_opt_fro})}, the objective is to minimize the Frobenius norm ${\lVert \mathbf{F}^{zf} - \mathbf{F}^{RF} \mathbf{F}^{BB}\rVert}_F$ with respect to the analog precoder $\mathbf{F}^{RF}$ while keeping the digital precoder $\mathbf{F}^{BB}$ fixed. This problem is initially solved without the unit norm constraint on $\mathbf{F}^{RF}$. Once $\mathbf{F}^{RF}$ is optimized, its elements are normalized as described by \textcolor{blue}{(\ref{eq:Update_RF})} to ensure that each element has a unit magnitude.
Algorithm~\textcolor{blue}{\ref{al:1}} shows the iterative procedure for solving this optimization problem. Steps 2 and 3 fix $\mathbf{F}^{BB}$ and update $\mathbf{F}^{RF}$. After updating $\mathbf{F}^{RF}$ in step 4, its elements are normalized. Subsequently, steps 5 and 6 fix $\mathbf{F}^{RF}$ and optimize $\mathbf{F}^{BB}$ by solving \textcolor{blue}{(\ref{eq:Update_RF})}, which aims to minimize ${\lVert \mathbf{F}^{zf} - \mathbf{F}^{RF} \mathbf{F}^{BB}\rVert}_F$ with respect to $\mathbf{F}^{BB}$. This procedure continues until convergence, as controlled by the criterion in step 7.
}
Consequently problem \textcolor{blue}{(\ref{eq:objective_4})} can be addressed iteratively. The convergence speed can be affected by the initial value of Algorithm~\textcolor{blue}{\ref{al:1}}. Appropriate initialization can lead to enhanced and quicker convergence.
\begin{align}
	\mathbf{F}^{RF} = \begin{bmatrix}
		\label{eq:Update_RF}
		\frac{{F}^{RF}_{1,1}}{|{F}^{RF}_{1,1}|} & \frac{{F}^{RF}_{1,2}}{|{F}^{RF}_{1,2}|} & \cdots & \frac{{F}^{RF}_{1,N_{RF}}}{|{F}^{RF}_{1,N_{RF}}|} \\
		\frac{{F}^{RF}_{2,1}}{|{F}^{RF}_{2,1}|} & \frac{{F}^{RF}_{2,2}}{|{F}^{RF}_{2,1}|} & \cdots & \frac{{F}^{RF}_{2,N_{RF}}}{|{F}^{RF}_{2,N_{RF}}|} \\
		\vdots                                  & \vdots                                  & \ddots & \vdots                                            \\
		\frac{{F}^{RF}_{N,1}}{|{F}^{RF}_{N,1}|} & \frac{{F}^{RF}_{N,2}}{|{F}^{RF}_{N,2}|} & \cdots & \frac{{F}^{RF}_{N,N_{RF}}}{|{F}^{RF}_{N,N_{RF}}|} \\
	\end{bmatrix}.
\end{align}
{  In \textcolor{blue}{(\ref{eq:initial})} and \textcolor{blue}{(\ref{eq:initial2})}, $\mathbf{\Sigma}$, $\mathbf{U}$, and $\mathbf{V}$ are matrices obtained from the Singular Value Decomposition (SVD) of the zero-forcing beamformer $\mathbf{F}^{zf}$, and the SVD is given by $\mathbf{F}^{zf} = \mathbf{U}\mathbf{\Sigma}\mathbf{V}^H$.
$\mathbf{\Sigma}$ is a diagonal matrix containing the singular values of $\mathbf{F}^{zf}$. $\mathbf{U}$ and $\mathbf{V}$ are unitary matrices representing the left and right singular vectors of $\mathbf{F}^{zf}$, respectively.}
The initial values of Algorithm~\textcolor{blue}{\ref{al:1}} are determined by selecting the first $N_{RF} = m$ largest singular values from $\mathbf{\Sigma}$, along with the corresponding first $N_{RF}$ columns of $\mathbf{U}$ and $\mathbf{V}$. This selection assumes that $N_{RF}$ equals the number of users, as shown below:
\begin{subequations}
	\begin{align}
		\label{eq:initial}
		\mathbf{F}^{RF}_m                    & = \mathbf{U}_m,                     \\
		\label{eq:initial2}	\mathbf{F}^{BB}_m & = \mathbf{\Sigma}_m \mathbf{V}_m^H.
	\end{align}
\end{subequations}
Utilizing this matrix approximation enhances the quality and speed of algorithm convergence. To iteratively solve \textcolor{blue}{(\ref{eq:part_1_opt_fro})} and \textcolor{blue}{(\ref{eq:part_2_opt_fro})}, any solver (e.g., the MATLAB Optimization toolbox, CVX) can be employed, with the steps illustrated in Algorithm~\textcolor{blue}{\ref{al:1}}. The convergence of the overall {  iterations between the zero-forcing fully digital beamformer and the hybrid beamformer, which consists of both analog and digital precoders} is depicted in Fig.~\ref{fig:convergence}.
\subsection{Problem of Spectrum Allocation}\label{sec:spec}
Once the hybrid beamforming matrix is determined, the optimization becomes solely dependent on bandwidth coefficients and backhaul bottleneck constraints. This simplifies the problem in \textcolor{blue}{(\ref{eq:objective})} as \textcolor{blue}{(\ref{eq:objective_after_bf})}. The optimization problem can be reformulated by substituting $\mu_{t}$ with $1-\mu_{a}$ and the objective expressed as $\left(\mu_a w\right) \sum_{j=1}^{J-1} R_{a, j}+{\left(1-\mu_a\right)w \sum_{i=1}^I R_{t, i}} = \left(\mu_a w\right) (\sum_{j=1}^{J-1}R_{a, j} - \sum_{i=1}^I R_{t, i}) +{ w \sum_{i=1}^I R_{t, i}}$. It is possible to omit the last term in this equation from the optimization problem, and constraint \textcolor{blue}{(\ref{eq:c_backhaul__after_bf})} can be reformulated accordingly:
\begin{subequations}
	\begin{align}
		\max_{\substack{\mu_{a}, \mu_{t}}}
		            & \label{eq:objective_after_bf} \quad \left(\mu_a\right) w\sum_{j=1}^{J-1} R_{a, j}+{\left(\mu_t\right) w \sum_{i=1}^I R_{t, i} } \\
		\text{s.t.} & \label{eq:ca__after_bf} \quad 0 \leq \mu_a \leq 1,                                                                                  \\
		            & \label{eq:ct__after_bf} \quad 0 \leq \mu_t \leq 1,                                                                                  \\
		            & \label{eq:sum_mu_a_t} \quad \mu_a + \mu_t = 1,                                                                                   \\
		            & \label{eq:c_backhaul__after_bf} \quad \mu_t\sum_{i= 1}^{I} {R_{t,i}} \leq \mu_a \mathrm{R_{a,b}}.
	\end{align}
\end{subequations}
 Thus, \textcolor{blue}{(\ref{eq:objective_after_bf})} can be represented as follows:
\begin{subequations}
	\begin{align}
		\max_{\substack{\mu_{a}}}
		            & \label{eq:objective_after_bf_omitted} \quad \left(\mu_a w\right) (\sum_{j=1}^{J-1}R_{a, j} - \sum_{i=1}^I R_{t, i})                    \\
		\text{s.t.} & \label{eq:ca__after_bf_omitted} \quad 0 \leq \mu_a \leq 1,                                                                                 \\
		            & \label{eq:c_backhaul__after_bf_omitted} \quad \mu_a \geq \frac{\mathrm{\sum_{i= 1}^{I}R_{t,i}}} {R_{a,b} + \sum_{i= 1}^{I} {R_{t,i}} }.
	\end{align}
\end{subequations}
{{Backhaul constraint \textcolor{blue}{(\ref{eq:c_backhaul__after_bf})} directly represents how the aerial backhaul link limits the rate of terrestrial users. By substituting $\mu_{t} = 1-\mu_{a}$, a lower bound for $\mu_{a}$ can be derived as \textcolor{blue}{(\ref{eq:c_backhaul__after_bf_omitted})}. This lower bound, combined with the upper bound in \textcolor{blue}{(\ref{eq:ca__after_bf_omitted})}, ensures that both aerial and terrestrial links always receive a non-zero bandwidth allocation.
On the other hand the closed-form solution in \textcolor{blue}{(\ref{eq:cf4})}, shows that \(\mu_a\) always takes the form of a fraction less than one.}  Problem \textcolor{blue}{(\ref{eq:objective_after_bf_omitted})} has a concave linear objective, and the only variable is the bandwidth coefficient. If the sum-SE of aerial users becomes greater than the sum-SE of terrestrial users in \textcolor{blue}{(\ref{eq:c_backhaul__after_bf_omitted})}, a positive value will be obtained for the concave objective, and the maximization will be satisfied if the bandwidth coefficient reaches its maximum value while adhering to the constraints. Furthermore, suppose the sum-SE of terrestrial users becomes larger than the sum-SE of aerial users. In this case, the objective takes a negative value, and the bandwidth coefficient must be as small as possible while satisfying the constraints \textcolor{blue}{(\ref{eq:ca__after_bf_omitted})} and \textcolor{blue}{(\ref{eq:c_backhaul__after_bf_omitted})} to maximize the objective function. Consequently, a closed-form expression can be derived as follows:
\begin{subequations}
	\begin{align}
		\label{eq:cf1} \hspace{40pt} \mathfrak{R}_t  = & \sum_{i= 1}^{I}R_{t,i},\hspace{10pt} \mathfrak{R}_a  =\sum_{j= 1}^{J}R_{a,j},                                                                                \\
		\label{eq:cf3} {\mu_a}^*                       & = \frac{\textrm{max}(\frac{\mathfrak{R}_t(\mathfrak{R}_a - \mathfrak{R}_t)}{\mathfrak{R}_t + R_b},\mathfrak{R}_a - \mathfrak{R}_t)}{\mathfrak{R}_a - \mathfrak{R}_t}, \\
		\label{eq:cf4}                                 & = \frac{\mathfrak{R}_t}{\mathfrak{R}_t + R_b}.
	\end{align}
\end{subequations}
Because of the equal power allocation in both the IAB-donor and IAB-node, along with the additional power required for the backhaul link, the SE of aerial users will be lower than that of terrestrial users, as shown in Fig.~\ref{fig:SE_t_a}. Thus, when the number of aerial and terrestrial users is assumed to be the same, \textcolor{blue}{(\ref{eq:cf3})} can be expressed as \textcolor{blue}{(\ref{eq:cf4})}.
\subsection{Non-IAB Scenario}\label{sec:non-IAB}
{  To comprehensively evaluate our proposed IAB-assisted UAV network, we performed a comparative analysis against a non-IAB scenario. In the non-IAB scenario, UAVs operate without IAB technology, leading to separate access and backhaul links. This separation requires dedicated resources, such as frequency spectrum, antennas, and hardware, making it impossible to share resources between the links. Consequently, the non-IAB scenario assumes a fixed bandwidth allocation for both access and backhaul links. In this context, the bandwidth coefficient remains fixed, regardless of the varying traffic demands of aerial or terrestrial users. Therefore, in non-IAB mode, the bandwidth coefficient is not considered an optimization variable in \textcolor{blue}{(\ref{eq:objective})}. Instead, only the hybrid beamforming matrices need to be optimized in \textcolor{blue}{(\ref{eq:objective_2})} and \textcolor{blue}{(\ref{eq:objective_3})} using Algorithm~\ref{al:1} to maximize the sum-rate for both aerial and terrestrial users. Numerical results demonstrate that the total sum-rate of the network in the non-IAB mode is lower than that in the IAB mode. Unlike the IAB mode, the non-IAB scenario cannot optimally utilize available resources in other links. Consequently, if a traffic overload occurs in the backhaul or access links, UAVs are unable to meet the requirements of terrestrial networks due to the lack of shared resources.}
\subsection{The Impact of Bandwidth on Thermal Noise}\label{sec:VII}
This section analyzes how bandwidth affects thermal noise in our mmWave system model. When we consider the thermal noise properties more accurately, the Johnson-Nyquist formula allows us to express $\sigma^2$, which is the noise power in watts, as a function of bandwidth like $\sigma^2(\mu w) = \mu w P_n$, where $w$ is bandwidth (in \textrm{Hz}), and $P_n$ is noise power spectral density (in \textrm{W}$/$\textrm{Hz}). To reformulate the optimization problem~\textcolor{blue}{(\ref{eq:objective_after_bf})} while taking into account the impact of noise-BW, we obtain the following problem:
\begin{subequations} 
	\begin{align}
		\max_{\substack{\mu_{a}                                      }}
		            & \label{eq:BW_objective_after_bf} \quad \left(\mu_a\right) w\sum_{j=1}^{J-1} R_{a, j}(\mu_a)+{\left(1-\mu_a\right) w \sum_{i=1}^I  R_{t, i}(1-\mu_a) } \\
		\text{s.t.} & \label{eq:BW_ca__after_bf} \quad 0 \leq \mu_a \leq 1,                                                                                                     \\
		            & \label{eq:BW_c_backhaul__after_bf} \quad (1-\mu_a)\sum_{i= 1}^{I} {R_{t,i}}(1-\mu_a) \leq \mu_a \mathrm{R_{a,b}}(\mu_a ).
	\end{align}
\end{subequations}
We can express the rates as follows for aerial users:
\begin{align}
	\label{eq:opt3_a}    R_{a,j}(\mu_a) & =  \log_{2} \left(1+\frac{\left|{g}_{a, j}\right|^2}{{\mathfrak{I}}_{a, j} +\mu_a w P_n}\right),
\end{align}
and for terrestrial users:
\begin{align}
	\label{eq:opt3_t}    R_{t,i}(\mu_a) & =  \log_{2} \left(1+\frac{\left|{g}_{t, i}\right|^2}{{\mathfrak{I}}_{t, i} +(1-\mu_a) w P_n}\right),
\end{align}
and for the backhaul link:
\begin{align}
	\label{eq:opt3_b}       R_{a, b}(\mu_a) & = \log_{2} \left(1+\frac{\left|{g}_{b}\right|^2}{{\mathfrak{I}}_{a,b} +(\mu_a) w P_n}\right).
\end{align}
\begin{algorithm}[t]
	\SetKwInOut{Input}{Input}
	\SetKwInOut{Output}{Output}
	\SetKwInOut{Return}{Return}
	\SetKwInOut{Initial}{Initial}
	\caption{Successive Convex Approximation (SCA) algorithm for problem~\textcolor{blue}{(\ref{eq:opt3_1})}}\label{al:2}
	\Input{$\mathbf{h}_{a, j}, \mathbf{h}_{t, i}, \mathbf{h}_{a, b},  \mathbf{F}^{RF}, \mathbf{F}^{BB}$ and  $P_n$}
	\Output{$\mu_a, \mu_t$}
	\Initial {$\mu^m_a$}
	Approximate objective and constraints by \textcolor{blue}{(\ref{eq:taylor1})} and \textcolor{blue}{(\ref{eq:taylor2})}\;
	\While{$|\mu^{m-1}_a-\mu^m_a| \geq \epsilon$}
		{$ \mu^*_a = arg~\max_{} \left(\mu_a\right) w\sum_{j=1}^{J-1} \hat{R}_{a, j}(\mu_a)+{\left(1-\mu_a\right) w \sum_{i=1}^I  \hat{R}_{t, i}(1-\mu_a) }$\;
		\text{s.t.} \textcolor{blue}{(\ref{eq:opt3_2})} and \textcolor{blue}{(\ref{eq:opt3_3})}\;
		$\mu^{m-1}_a \gets \mu^m_a $\;
		$\mu^m_a \gets \mu^*_a $;
		}
\end{algorithm}
This problem is non-convex because the second derivative of the objective depends on channel coefficients, which may be convex or concave depending on channel coefficient values. $\left|{g}_{a, j}\right|^2$, $\left|{g}_{t, i}\right|^2$ and $\left|{g}_{b}\right|^2$ represent the received power at aerial UEs, terrestrial UEs and backhaul node respectively, whereas $\mathfrak{I}_{a, j}$, $\mathfrak{I}_{t, i}$ and $\mathfrak{I}_{a,b}$ denote the interference power. We try to rewrite this problem and then use an SCA-based algorithm to cope with this non-convexity. We can reformulate \textcolor{blue}{(\ref{eq:opt3_a})}, \textcolor{blue}{(\ref{eq:opt3_t})} and \textcolor{blue}{(\ref{eq:opt3_b})} as follows:
\begin{subequations}
	\begin{align}
		R_{a,j}(\mu_a) & = \log_2 \left(\frac{{\tilde{g}}_{a, j} + \mu_a w P_n}{{\mathfrak{I}}_{a, j} +\mu_a w P_n}\right) \label{eq:SE1}         \\
		               & = \log_2 \left(\tilde{g}_{a, j} + \mu_a w P_n \right) - \log_2 \left({{\mathfrak{I}}_{a, j} +\mu_a w P_n}\right),  \notag \\
		R_{t,i}(\mu_a) & = \log_2 \left(\frac{{\tilde{g}}_{t, i} + (1-\mu_a) w P_n}{{\mathfrak{I}}_{t, i} +(1-\mu_a) w P_n}\right)\label{eq:SE2} \\
		               & = \log_2 \left(\tilde{g}_{t, i} + (1-\mu_a) w P_n \right)  \notag                                                        \\
		               & \quad\quad - \log_2 \left({{\mathfrak{I}}_{t, i} +(1-\mu_a) w P_n}\right),   \notag                                       \\
		R_{a,b}(\mu_a) & = \log_2 \left(\frac{{\tilde{g}}_{a, b} + \mu_a w P_n}{{\mathfrak{I}}_{a, b} +\mu_a w P_n}\right) \label{eq:SE3}         \\
		               & = \log_2 \left(\tilde{g}_{a, b} + \mu_a w P_n \right) - \log_2 \left({{\mathfrak{I}}_{a, b} +\mu_a w P_n}\right). \notag
	\end{align}
\end{subequations}
An approximation of aerial rates can be found using a first-order Taylor series as follows:
\begin{subequations}
	\begin{align}
		 & \log_{2} \left(\tilde{g}_{a, j} + \mu_a w P_n \right)  \approx                                                                         \\
		 & \log_{2} \left(\tilde{g}_{a, j} + \mu^m_a w P_n \right) + \frac{w P_n(\mu_a - \mu^m_a)}{ln(2)(\tilde{g}_{a, j} + \mu^m_a w P_n)}
		\label{eq:taylor1} = L_1(\mu_a), \notag                                                                                                    \\
		 & \log_{2} \left(\mathfrak{I}_{a, j} +\mu_a w P_n \right)  \approx                                                                       \\
		 & \log_{2} \left(\mathfrak{I}_{a, j} + \mu^m_a w P_n \right) + \frac{w P_n(\mu_a - \mu^m_a)}{ln(2)(\mathfrak{I}_{a, j} + \mu^m_a w P_n)}
		\label{eq:taylor2}  = L_2(\mu_a), \notag
	\end{align}
\end{subequations}
and then:
\begin{subequations}
	\begin{align}
		 & R_{a,j}(\mu_a)  \approx L_1(\mu_a) - L_2(\mu_a) = \hat{R}_{a,j}.
	\end{align}
\end{subequations}
In \textcolor{blue}{(\ref{eq:taylor1})} and \textcolor{blue}{(\ref{eq:taylor2})}, $\mu^m_a$ represents the value of $\mu_a$ in the ${m}^{th}$ iteration of the SCA algorithm. Additionally, we have these approximations for terrestrial and backhaul links denoted as $\hat{R}_{t,i}$ and $\hat{R}_{a,b}$. Therefore, problem~\textcolor{blue}{(\ref{eq:BW_objective_after_bf})} can be expressed as follows:
\begin{subequations}
	\begin{align}
		\max_{\substack{\mu_{a}                                    }}
		            & \label{eq:opt3_1} \quad \left(\mu_a\right) w\sum_{j=1}^{J-1} \hat{R}_{a, j}(\mu_a)+{\left(1-\mu_a\right) w \sum_{i=1}^I  \hat{R}_{t, i}(1-\mu_a) } \\
		\text{s.t.} & \label{eq:opt3_2} \quad 0 \leq \mu_a \leq 1,                                                                                                           \\
		            & \label{eq:opt3_3} \quad (1-\mu_a)\sum_{i= 1}^{I} {\hat{R}_{t,i}}(1-\mu_a) \leq \mu_a \mathrm{\hat{R}_{a,b}}(\mu_a ).
	\end{align}
\end{subequations}

The SE can be represented as a linear function of the bandwidth coefficient $\mu_a$. The resulting quadratic problem can be solved using Algorithm~\textcolor{blue}{\ref{al:2}} \textcolor{blue}{\cite{li2022exploring}},  \textcolor{blue}{\cite{xie2022sum}}.
\section{Simulation Results}\label{sec:IV}
This section provides a numerical analysis. The parameters utilized in the simulation are listed in Table~\ref{tb:table_params}. {  First, the convergence of the proposed algorithms is demonstrated and validated. Subsequently, the proposed hybrid beamforming method is compared with other hybrid beamforming and fully digital methods. The superiority of the theoretical analysis in terms of sum-rate maximization is compared with the traditional non-IAB scenario. The closed-form solution is presented as a function of power and the number of antennas through various simulations. Additionally, the proposed method is analyzed in terms of energy efficiency. A contour plot illustrating the simultaneous variation in both power and bandwidth and its effect on the network's sum-rate is provided, and the results are compared with the non-IAB scenario.}
\begin{table}
	\renewcommand{\arraystretch}{0.97}
	\caption{Simulation parameters}
	\label{tb:table_params}
	\centering
	\begin{tabular}{|c|c|c|}
		\hline
		\bfseries Parameter         & \bfseries Notation  & \bfseries Value                      \\
		\hline
		Frequency                   & $f_c$               & $28~$\textrm{GHz}                             \\
		IAB-donor height                & $\mathrm{h_{uav}}$  & $150~$\textrm{m}                               \\
		IAB-node height                & $\mathrm{h_{cuav}}$ & $70~$\textrm{m}                              \\
		Aerial UEs height           & $\mathrm{h_{auav}}$ & $100~$\textrm{m}                              \\
		Aerial links      &$J$        & $4$                                    \\
		Terrestrial links & $I$        & $3$                                  \\
		Bandwidth                   & $w$                 & $500~$\textrm{MHz}                            \\
		PSD of thermal noise        & $\mathrm{P_n}$      & $-170~$\textrm{dBm$/$Hz} \textcolor{blue}{\cite{mozaffari20183d}} \\
		RF chains power             & $\mathrm{P}_{RF}$   & $250~$\textrm{mW}   \textcolor{blue}{\cite{gao2016energy}}      \\
		Phase shifters power        & $\mathrm{P}_{PS}$   & $1~$\textrm{mW}  \textcolor{blue}{\cite{gao2016energy}}         \\
		Path loss exponent          & $n$                 & $2$                                  \\
  		Reference path loss          & $\gamma_{0}$              & $-61.39$ \textrm{dB}                           \\

		RF chains                   & $N_{RF}$            & 3                                    \\
		Antenna elements            & $N$                 & $64$                                 \\
		Antenna dimensions          & $N_x \times N_y$    & $8\times8$                           \\
		\hline
	\end{tabular}
\end{table}
\subsection{Algorithms Convergence and Analysis of Proposed Hybrid Beamforming Technique}
{  Algorithm~\ref{al:1} and Algorithm~\ref{al:2} are separate and independent algorithms in our proposed solution. Algorithm~\ref{al:1} focuses on the hybrid beamforming design problem, while Algorithm~\ref{al:2}, discussed in Sec.~\ref{sec:III}, addresses the successive convex approximation for the bandwidth allocation problem under the bandwidth-dependent thermal noise assumptions. Without assuming the bandwidth-dependent thermal noise effect, we use the derived closed-form solution in \textcolor{blue}{(\ref{eq:cf4})} to find the bandwidth allocation coefficient instead of Algorithm~\ref{al:2} and these two algorithms are not part of a joint iterative process.}
\begin{figure}[tp]
	\centering
	\includegraphics[width=0.45\textwidth]{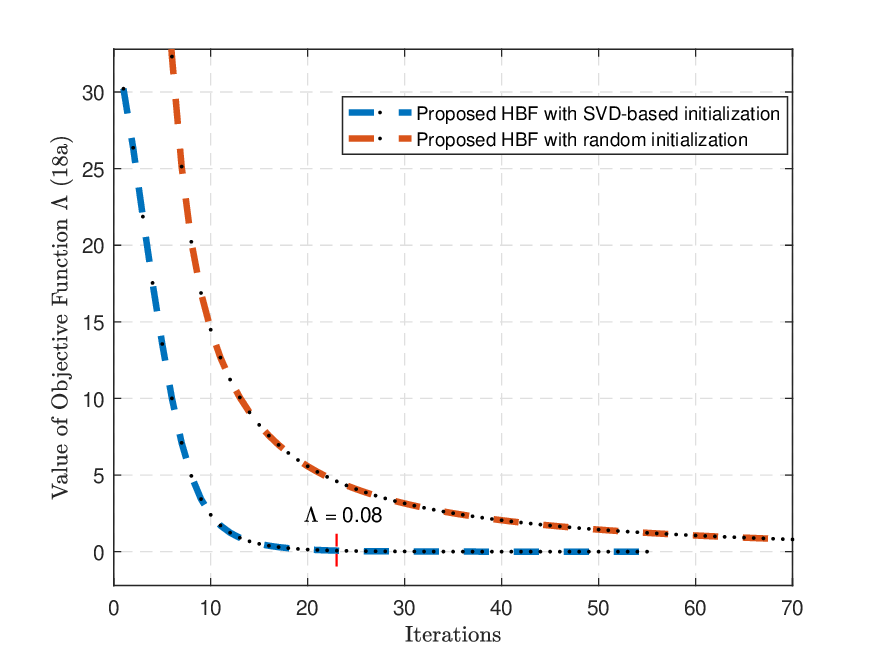}
	\caption{Convergence of hybrid beamforming (HBF) Algorithm~\ref{al:1} based on the different initialization methods.}
	\label{fig:convergence}
\end{figure}
\begin{figure}[tp]
	\centering
	\includegraphics[width=0.45\textwidth]{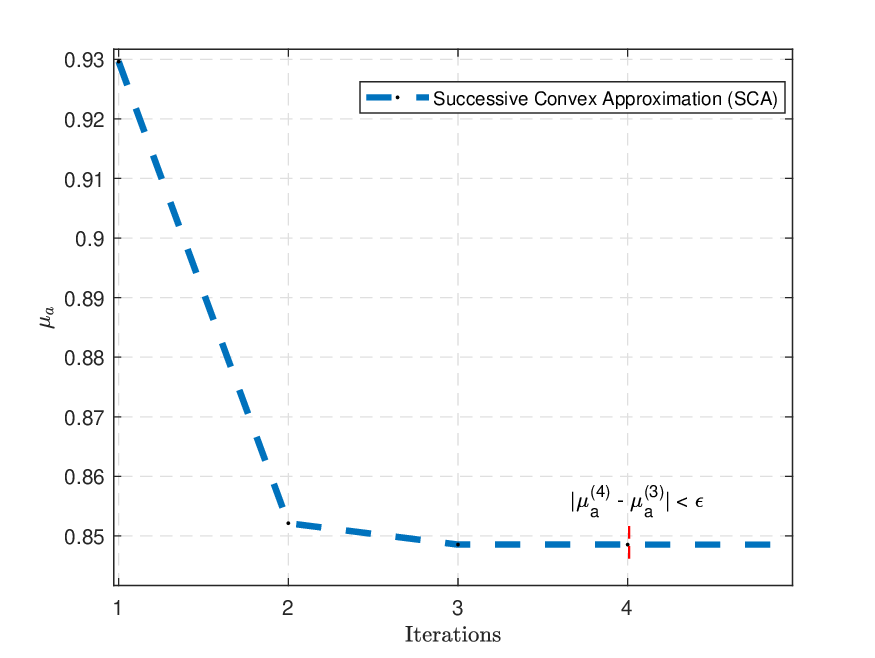}
	\caption{Convergence of Successive Convex Approximation (SCA) Algorithm~\ref{al:2}.}
	\label{fig:convergence2}
\end{figure}
Fig.~\ref{fig:convergence} illustrates the convergence of the hybrid beamforming algorithm, which is achieved after $23$ iterations. { With the SVD-initialization performed in the proposed Algorithm~\ref{al:1}, the convergence speed is notably faster than that of the random-initialization algorithm, as referenced in \textcolor{blue}{\cite{majidzadeh2017partially}.} }The number of iterations required can be limited, or the algorithm's stopping condition may be based on the value of the objective function. In the conducted simulation, Algorithm~\ref{al:1} stops upon reaching a specific target value ($\Lambda \leq 0.1$). However, should there be changes in the channel conditions, the algorithm must be rerun to determine the hybrid beamforming matrices, comprising both analog and digital precoders. {  Fig.~\ref{fig:convergence2} shows the convergence of  the SCA Algorithm~\ref{al:2} used to solve the optimization problem~\textcolor{blue}{(\ref{eq:opt3_1})} considering the impact of bandwidth on thermal noise. The y-axis represents the value of the aerial BW coefficient $\mu_a$. The plot demonstrates that after approximately 4 iterations, the difference between the values of $\mu_a$ in consecutive iterations becomes smaller than the predefined criterion $\epsilon = 0.001$. This convergence criterion is used as the stopping condition for the SCA algorithm.
}

\begin{figure}[tp]
	\centering
	\includegraphics[width=0.5\textwidth]{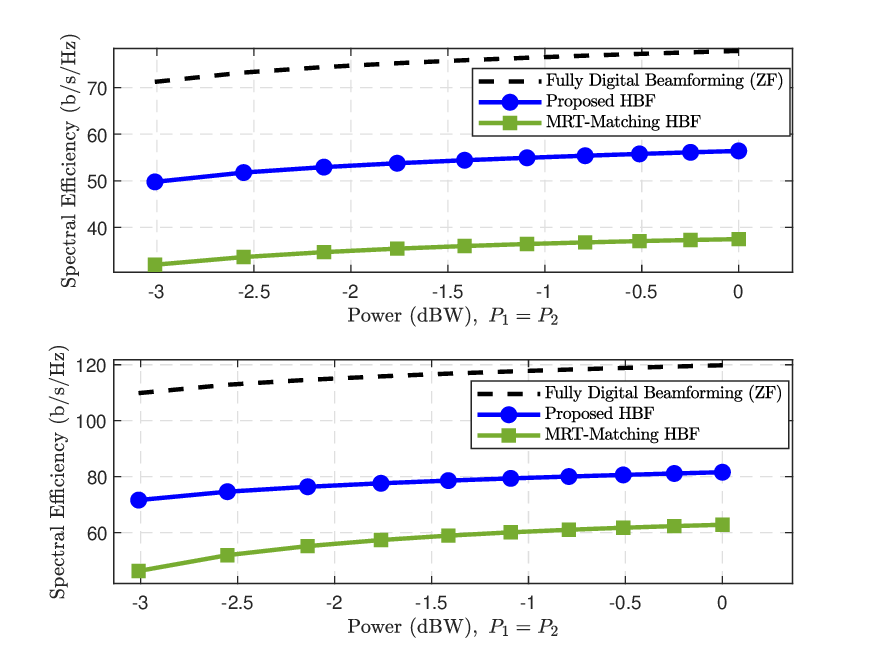}
	\caption{Sum Spectral Efficiency (SE) comparison of the proposed hybrid beamforming (HBF) technique against different beamforming methods for aerial UEs (top) and terrestrial UEs (bottom), considering variable transmit power at the IAB-donor and IAB-node with a fixed number of UEs ($I = 3, J = 4$).}
	\label{fig:SE_t_a}
\end{figure}
{  In Fig.~\ref{fig:SE_t_a}, we compare our proposed Hybrid Beamforming (HBF) with Maximum Ratio Transmission (MRT)-matching HBF (based on \textcolor{blue}{\cite{majidzadeh2017partially}}) and fully digital beamforming (zero-forcing) methods. As shown, our proposed HBF performs 41.32\% better on average than MRT-matching in terms of achieved sum-SE, in addition to having a faster convergence speed as previously discussed.} Equal power allocation is assumed for both the IAB-node and IAB-donor and the power of the backhaul link is also equal to the power required by terrestrial users. Consequently, the SE of aerial users is expected to be lower than that of terrestrial users, as seen in Fig.~\ref{fig:SE_t_a}. However, hybrid beamforming is less effective than fully digital beamforming, which is particularly evident in the MISO scenario where hybrid beamforming underperforms compared to the MIMO scenario. Additionally, the hybrid beamforming configuration has fewer RF chains than fully digital beamforming, where the number of RF chains equals the number of sub-arrays elements. For hybrid beamforming, three RF chains are used, whereas fully digital beamforming requires 64 RF chains. Hence, the fully digital approach leads to significant issues in energy consumption and hardware size, especially in UAVs with limited hardware and energy storage. Therefore, the 39.4\% drop in SE for terrestrial users and the 48.2\% drop for aerial users due to hybrid beamforming in the MISO scenario is reasonable from an energy standpoint.
\begin{figure}[tp]
	\centering
	\includegraphics[width=0.45\textwidth]{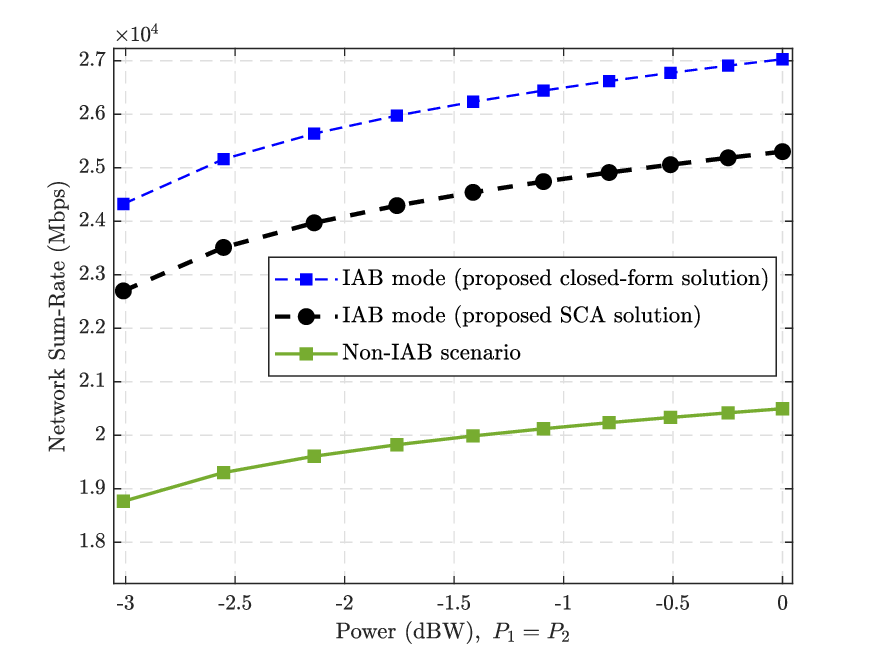}
	\caption{Sum-rate comparison of proposed IAB mode with dynamic bandwidth allocation (blue curve), considering bandwidth-dependent thermal noise (black curve), and non-IAB scenario with fixed bandwidth (green curve). This comparison considers variable power for the IAB-donor and IAB-nodes, with a constant number of UE ($I = 3, J = 4$).}
	\label{fig:sumrate_mu}
\end{figure}
\subsection{Outperformance Analysis of the Proposed IAB-assisted UAV Network}
{  Fig.~\ref{fig:sumrate_mu} presents a comparison of the sum-rate performance achieved under various network scenarios. The closed-form formula curve represents the theoretical maximum sum-rate obtained by solving the optimization problem \textcolor{blue}{(\ref{eq:objective_after_bf_omitted})} using the derived closed-form solution in \textcolor{blue}{(\ref{eq:cf4})}. This solution optimally allocates the bandwidth between aerial and terrestrial links to maximize the overall sum-rate. To validate the accuracy of the closed-form expression, we benchmark it against the Genetic Algorithm (GA) curve, which depicts the sum-rate achieved by numerically solving \textcolor{blue}{(\ref{eq:objective_after_bf})} using a genetic algorithm. As evident from Fig.~\ref{fig:Compare_GA}, the closed-form solution closely matches the GA benchmark, confirming its validity.
Furthermore, we investigate the performance of a traditional non-IAB network scenario, represented by the non-IAB scenario or fixed bandwidth allocation curve. In this case, as explained in Sec.~\ref{sec:non-IAB} it is impossible to dynamically optimize bandwidth and {its allocation between aerial and terrestrial links is fixed and equally divided, set at 0.5.} Notably, the sum-rate in this non-IAB scenario exhibits a significant drop compared to the optimized IAB scenarios, highlighting the performance gains achieved through the proposed IAB-assisted UAV network framework and optimized bandwidth allocation.
Additionally, we analyze the impact of considering the bandwidth-dependent thermal noise on the system performance. The SCA curve illustrates the sum-rate achieved when solving the optimization problem \textcolor{blue}{(\ref{eq:opt3_1})} using the SCA Algorithm~\textcolor{blue}{\ref{al:2}}. This approach accounts for the dependence of thermal noise power on the allocated bandwidth, providing a more realistic representation of our proposed framework. As expected, incorporating the bandwidth effect on noise leads to a further reduction in the achievable sum-rate compared to the scenario without considering this factor.}
\begin{figure}[tp]
	\centering
	\includegraphics[width=0.5\textwidth]{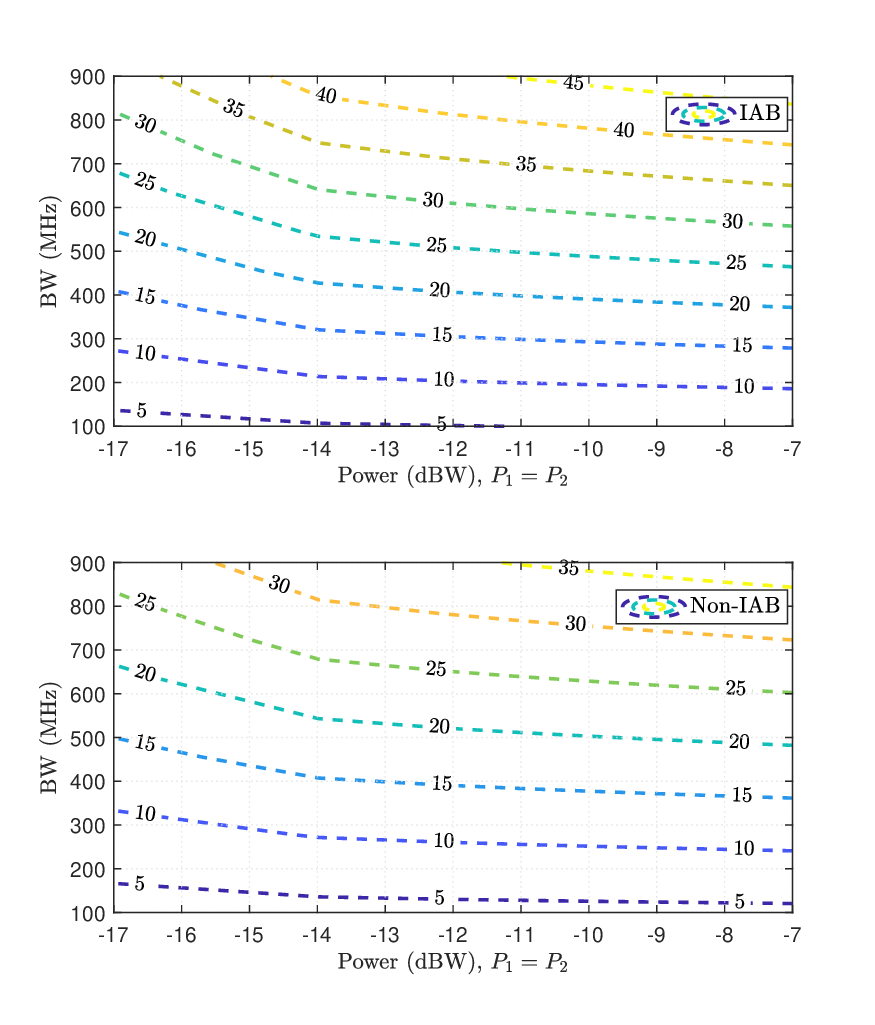}
	\caption{Contour plots of network sum-rate versus available bandwidth and transmit power for IAB mode (top) and non-IAB scenario (bottom).}
	\label{fig:contour1}
\end{figure}
Fig.~\ref{fig:contour1} demonstrates contour diagrams that illustrate the variation of the network's total sum-rate concerning the available bandwidth and the total available power ($P_1$ and $P_2$) at the IAB-donor and IAB-node, respectively. The two contour diagrams depict the sum-rate performance in the IAB mode (top figure) and the non-IAB mode (bottom figure). Each contour line represents a specific sum-rate value in Gbps.
{  In the IAB mode (top figure), the sum-rate contours are more densely packed, indicating a higher rate of change in the sum-rate as the available bandwidth and power vary. This behavior suggests that the IAB mode can effectively leverage the available resources to achieve higher sum-rates. For instance, to attain a sum-rate of $25~$\textrm{Gbps}, the IAB mode requires approximately $500~$\textrm{MHz} of bandwidth and an available power of $-12~$\textrm{dBW}, as indicated by the corresponding contour line.
On the other hand, in the non-IAB mode (bottom figure), the contour lines are more widely spaced, indicating a slower rate of change in the sum-rate with respect to the available bandwidth and power. This behavior suggests that the non-IAB mode is less efficient in utilizing the available resources. To achieve the same sum-rate of $25~$\textrm{Gbps} in the non-IAB mode, a higher bandwidth of approximately $650~$\textrm{MHz} is required, even with an available power of $-12~$\textrm{dBW}.}
\begin{figure}[tp]
	\centering
	\includegraphics[width=0.45\textwidth]{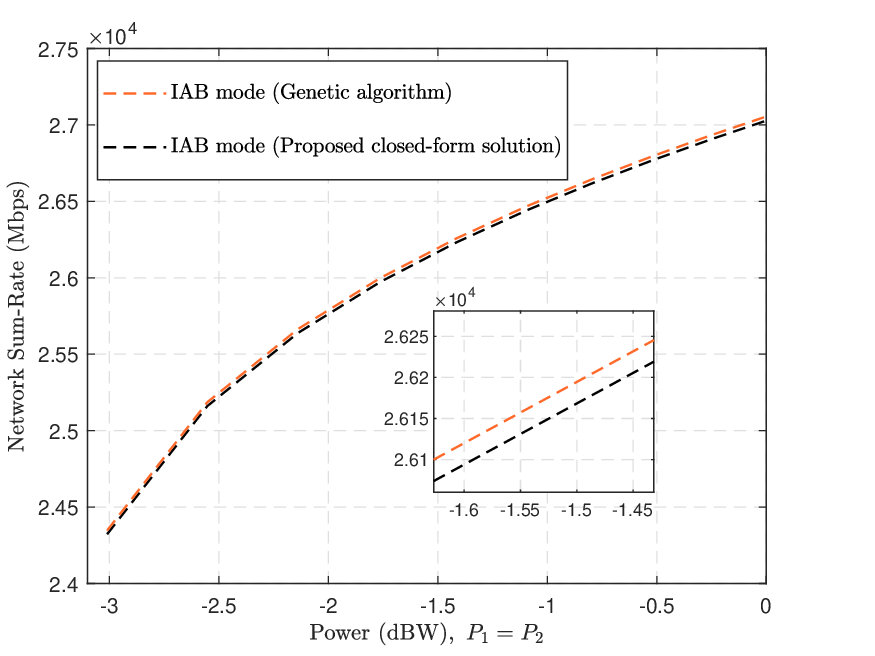}
	\caption{Comparison between the proposed closed-form solution and the genetic algorithm with a constant number of UE ($I = 3, J = 4$) }
	\label{fig:Compare_GA}
\end{figure}
\begin{figure}[tp]
	\centering
	\includegraphics[width=0.45\textwidth]{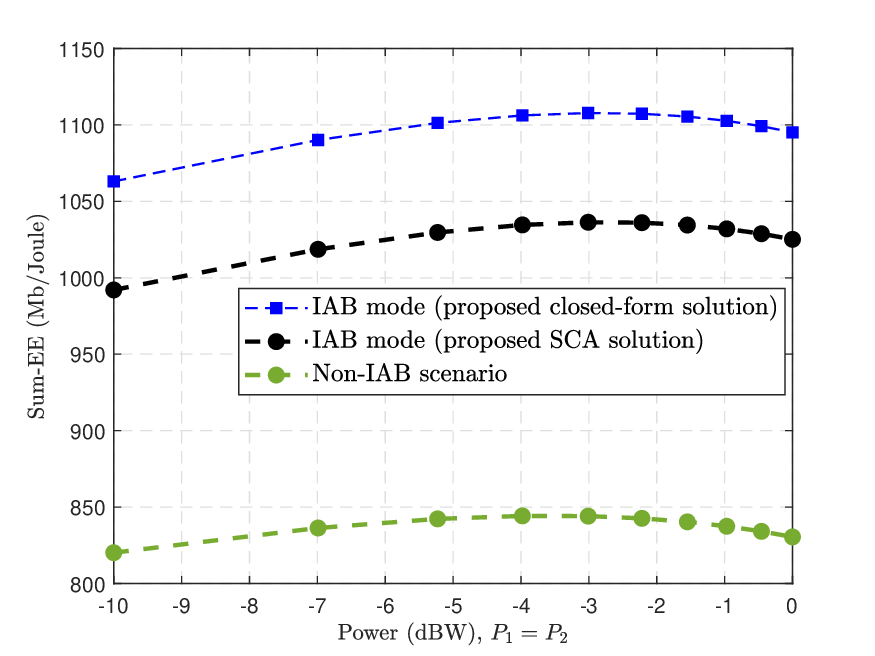}
	\caption{Sum Energy Efficiency (Sum-EE) comparison of proposed IAB mode with dynamic bandwidth allocation (blue curve), considering bandwidth-dependent thermal noise (black curve), and non-IAB scenario with fixed bandwidth (green curve). This comparison considers variable power for the IAB-donor and IAB-nodes, with a constant number of UE ($I = 3, J = 4$), the number of transmit antennas $N = 64$, and the assumed number of RF chains is $N_{RF} = 3$.}
	\label{fig:ee_comparison}
\end{figure}
This comparison between the two contour diagrams highlights the superior performance of the proposed IAB mode over the non-IAB mode in terms of sum-rate and resource utilization. The IAB mode can achieve higher sum-rates while consuming less resources, making it more efficient and suitable for aerial wireless backhaul scenarios.
Furthermore, the contour diagrams provide valuable insights into the trade-offs between bandwidth and power in achieving desired sum-rate targets. For instance, in the IAB mode, if the available bandwidth is limited, a higher available power can compensate to some extent and still achieve a reasonable sum-rate. Conversely, if the available power is constrained, increasing the bandwidth allocation can help maintain the desired sum-rate performance.
Fig.~\ref{fig:ee_comparison} illustrates the variation of Sum-Energy Efficiency (Sum-EE) to the total available transmit power at the IAB-donor and IAB-node. The Sum-EE is calculated as the ratio of the total network sum-rate to the total power consumption, considering the transmit power and power consumption of Radio Frequency (RF) chains and phase shifters.
As the figure shows, the Sum-EE initially increases with an increment in the available transmit power for all three scenarios. This behavior is expected, as higher transmit power generally leads to improved signal-to-noise ratios and higher achievable data rates, thereby increasing the Sum-EE. However, beyond a certain power threshold (approximately $-3~$\textrm{dBW}), the Sum-EE starts to decrease for all scenarios.
This decrease in Sum-EE at higher power levels can be attributed to the sum-rate saturation effect. 
Notably, the proposed IAB mode outperforms the non-IAB scenario in terms of Sum-EE across the entire range of transmit power values considered. This superior performance can be related to the efficient utilization of available resources and ability to the optimal allocation of bandwidth between aerial and terrestrial links in the IAB mode. By dynamically adjusting the bandwidth allocation, the IAB mode can effectively adapt to varying traffic demands, ultimately achieving higher sum-rates and better energy efficiency.
{Fig.~\ref{fig:BW_coeff_comparison} illustrates the variation of the bandwidth allocation coefficient ($\mu_a$) as a function of the available power budget at the IAB-donor and IAB-node. The graph depicts the trend of $\mu_a$, which represents the fraction of the total bandwidth allocated to aerial links, including both access links for aerial users and the backhaul link.
\begin{figure}[tp]
	\centering
	\includegraphics[width=0.45\textwidth]{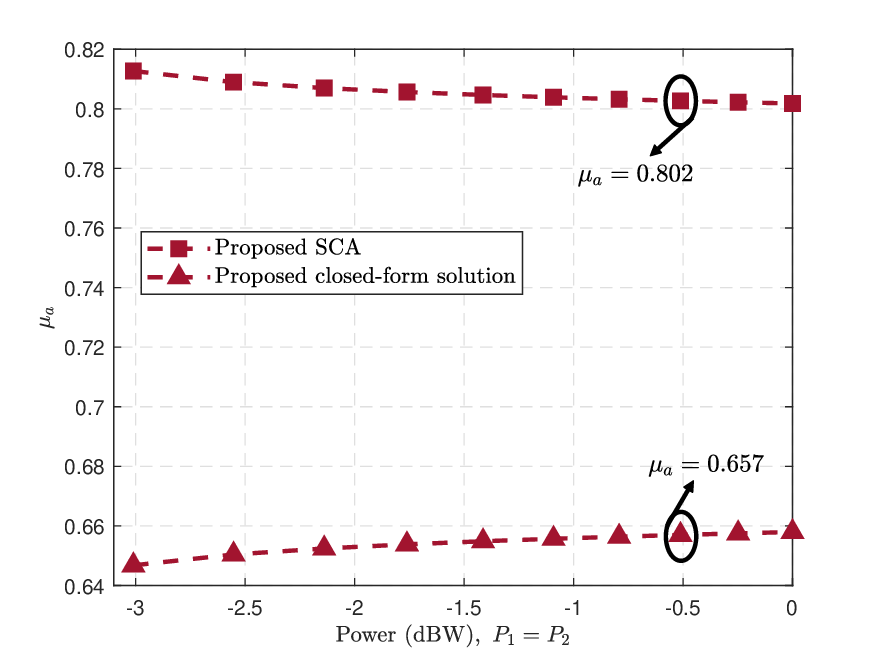}
	\caption{Variation of the optimal bandwidth allocation coefficient ($\mu_a$) as a function of the total power budget at the IAB-donor and IAB-node, considering both the closed-form solution and the scenario accounting for bandwidth-dependent thermal noise.}
	\label{fig:BW_coeff_comparison}
\end{figure}
\begin{figure}[tp]
	\centering
	\includegraphics[width=0.45\textwidth]{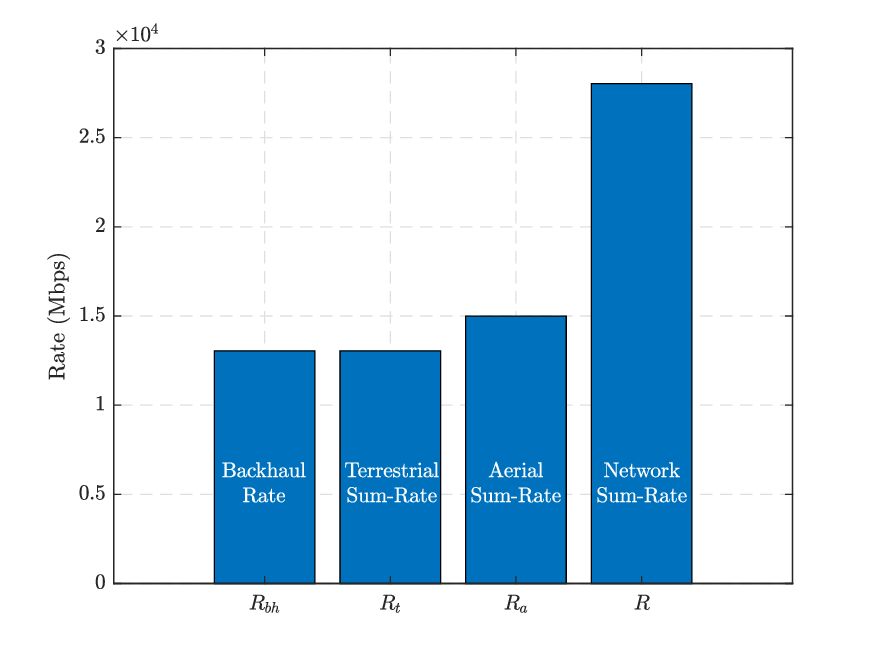}
	\caption{Illustration of backhaul bottleneck constraint satisfaction with sum-rates of backhaul link, terrestrial users, and aerial users.}
	\label{fig:compare_sumrate}
\end{figure}
\begin{figure*}[tp]
	\centering
	\includegraphics[width=0.9\textwidth]{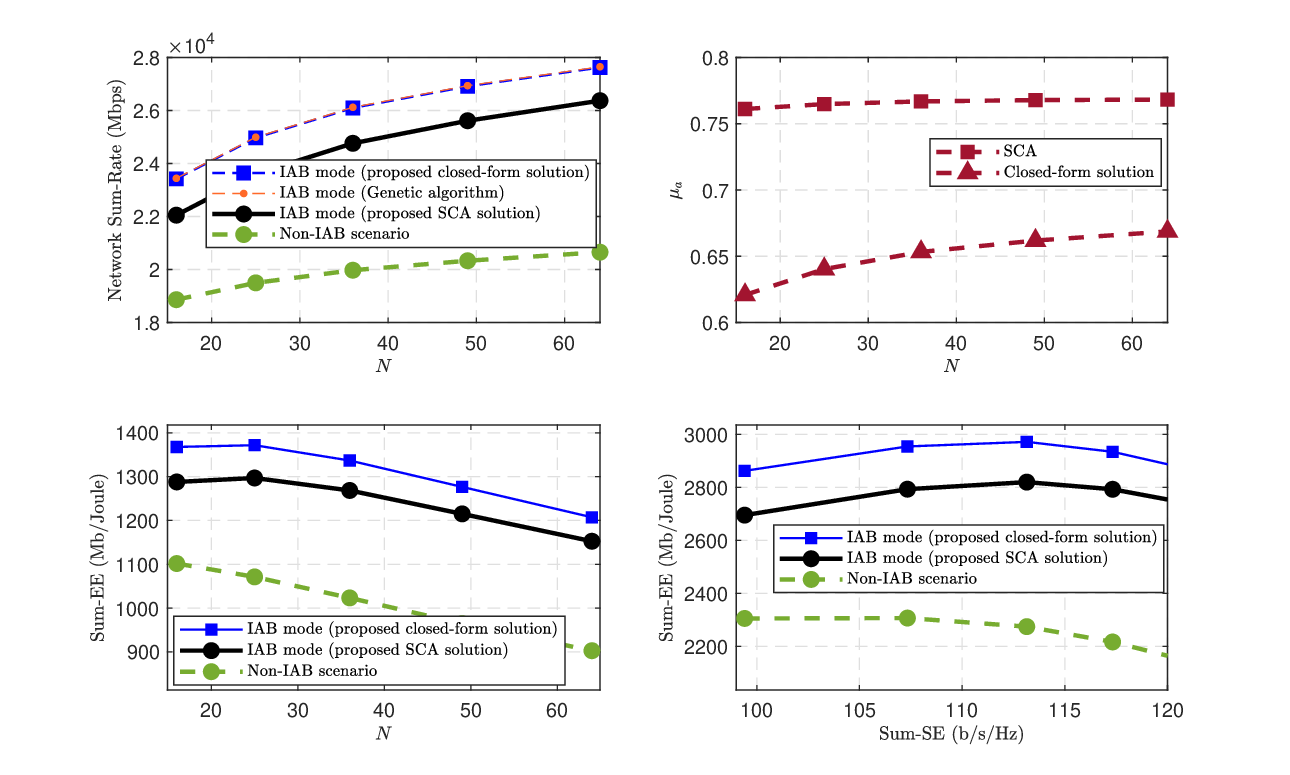}
	\caption{
 Analysis of the proposed IAB-assisted UAV framework versus number of transmit antennas (N) in terms of: sum-rate, bandwidth allocation coefficient ($\mu_a$), sum energy efficiency (Sum-EE), and Sum-EE vs sum spectral efficiency (Sum-SE) trade-off for IAB and non-IAB scenarios. It is assumed that $P_{tr} = 0$~\textrm{dBW} and $N_{RF} = 3$ and a constant number of UE ($I = 3, J = 4$).
 }
	\label{fig:antenna_sim}
\end{figure*}
As observed in the figure, in both scenarios of considering the bandwidth-dependent thermal noise solved by the SCA algorithm or using the closed-form solution, the value of $\mu_a$ is optimized as the available power changes. Its trend can be attributed to the SE formulation of aerial and terrestrial links, as expressed in \textcolor{blue}{(\ref{eq:SE1})} and \textcolor{blue}{(\ref{eq:SE2})} for bandwidth-dependent thermal noise, and in \textcolor{blue}{(\ref{eq:thr_a})} and \textcolor{blue}{(\ref{eq:thr_t})} for the closed-form solution. As discussed in Fig.~\ref{fig:SE_t_a}, the SE of aerial links is lower compared to terrestrial links, necessitating a higher bandwidth allocation to aerial links (higher $\mu_a$) to compensate for the lower SE. However, as the available power increases, the SE of aerial links improves, allowing for an optimized bandwidth allocation to aerial links while still maintaining the required data rates. The $\mu_a$ is optimized subject to the backhaul bottleneck constraint, as defined in equation \textcolor{blue}{(\ref{eq:c_backhaul__after_bf})}. This constraint ensures that the sum-rate of terrestrial users does not exceed the rate of the aerial backhaul link, which acts as a bottleneck. As the traffic load of terrestrial users grows, the bandwidth coefficient for aerial links ($\mu_a$) must be adjusted to accommodate the increasing backhaul requirements, ensuring that the backhaul link can support the aggregate data rate demanded by terrestrial users.
Furthermore, the graph demonstrates that the value of $\mu_a$ converges to a specific value as the available power increases. This convergence point corresponds to the scenario where the SE of aerial links becomes sufficiently high, and the bandwidth allocation is primarily governed by the backhaul bottleneck constraint rather than the individual link SEs}. The bar chart in Fig.~\ref{fig:compare_sumrate} illustrates the satisfaction of the backhaul bottleneck constraint \textcolor{blue}{(\ref{eq:c_backhaul__after_bf_omitted})}, ensuring that the total rate received by terrestrial users does not exceed the received rate at the IAB-node, which acts as the bottleneck link.

{ This constraint is a crucial aspect of the proposed IAB-assisted UAV network framework, as it ensures that the backhaul link can support the total data-rate demanded by terrestrial users. This bar chart presents network sum-rate ($R$) plus three distinct bars ($R_{bh}$, $R_{t}$, and $R_{a}$), each representing a different rate category. The first bar depicts the rate of the backhaul link, which serves as the bottleneck for the terrestrial users' rate. This bar has an equal height to the second bar, representing the sum-rate of terrestrial users. This equality demonstrates that the proposed optimization framework successfully allocates the bandwidth between aerial and terrestrial links, such that the backhaul rate precisely matches the total sum-rate received by terrestrial users. By adhering to the backhaul bottleneck constraint, the proposed framework ensures that the terrestrial network's performance is not limited by the backhaul capacity. This constraint is particularly important in scenarios where the traffic load generated by terrestrial users is high, as it guarantees that the backhaul link can accommodate the required rates without becoming a performance bottleneck.} Furthermore, it provides insights into the bandwidth allocation strategy adopted by the proposed framework. The higher sum-rate of aerial users, the third bar, compared to terrestrial users, suggests that a larger portion of the available bandwidth is allocated to aerial links, including the backhaul link. This allocation is necessary to accommodate the additional traffic load generated by terrestrial users, as well as the resource requirements of the backhaul link itself.
By following the aerial backhaul bottleneck constraint and optimizing the bandwidth allocation, the proposed IAB framework ensures efficient resource utilization and maximizes the overall network sum-rate, while simultaneously providing reliable backhaul connectivity for terrestrial users.

{  Fig.~\ref{fig:antenna_sim} illustrates the variation of different performance metrics, including the bandwidth allocation coefficient ($\mu_a$), the network's total sum-rate, and the sum-EE, as a function of the number of transmit antennas ($N$) employed at the IAB-donor and IAB-node. EE is computed using the formula $EE = \frac{Rate}{P_t}$, where $P_t = P_{tr} + N_{RF}P_{RF} + N_{PS}P_{PS}$, as in \textcolor{blue}{\cite{gao2016energy}} and \textcolor{blue}{\cite{roth2017achievable}}.

The top-left subplot depicts the network's total sum-rate as a function of the number of transmit antennas. As expected, the sum-rate increases with an increasing number of antennas for both the IAB and non-IAB scenarios. However, the proposed IAB-assisted UAV network consistently outperforms the non-IAB scenario across the entire range of antenna numbers considered. This superior performance can be attributed to the efficient utilization of available resources and the ability to optimally allocate the bandwidth between aerial and terrestrial links in the IAB mode which is not possible in non-IAB scenario.

The top-right subplot shows the variation of optimized $\mu_a$ with respect to the number of transmit antennas. It is important to note that the bandwidth allocation is optimized subject to the aerial backhaul bottleneck constraint, ensuring that the summation rate demanded by terrestrial users does not exceed the capacity of the aerial backhaul link.
The bottom-left subplot illustrates the variation of the Sum-EE with respect to the number of transmit antennas. The Sum-EE is calculated as the ratio of the total network sum-rate to the total power consumption, which includes the transmit power as well as the power consumption of RF chains and phase shifters. Similar to the sum-rate behavior, the Sum-EE increases with an increasing number of antennas for both scenarios, and the proposed IAB mode exhibits better energy efficiency compared to the non-IAB scenario. The bottom-right subplot illustrates the sum-EE vs. sum-SE plot. It is evident that in all ranges of sum-SE, the Sum-EE in IAB mode performs better than the non-IAB scenario

Overall, the results presented in Fig.~\ref{fig:antenna_sim} highlight the notable performance gains achieved by the proposed IAB-assisted UAV network framework in terms of sum-rate and energy efficiency. By leveraging the IAB technique and optimizing the bandwidth allocation between aerial and terrestrial links, the proposed framework can effectively utilize the available resources and adapt to varying traffic demands, leading to improved network performance. These findings underscore the suitability of the proposed IAB-assisted UAV network for future wireless communication systems that rely heavily on non-terrestrial networks.}
\section{Conclusion}\label{sec:V}
In this paper, we proposed an IAB-assisted UAV model for providing simultaneous aerial backhauling and access to terrestrial and aerial users. We employ the hybrid beamforming technique to mitigate interference between the aerial user's access links and the backhaul link, which share the same frequency band. Furthermore, we conduct a comprehensive analysis to optimize bandwidth allocation between aerial and terrestrial users, ultimately leading to a higher overall network sum-rate. Additionally, we investigate the impact of bandwidth on thermal noise and its influence on network performance. In summary, while optical fiber is well-suited for terrestrial communications, the wireless UAV-based communication scenario benefits from mmWave utilization and the IAB multi-hop architecture, significantly improving network performance. {  In future work, it is crucial to explore the integration of terrestrial and non-terrestrial networks in more realistic scenarios based on aerial mobile-IAB. This integration can account for factors such as imperfect channel conditions, UE mobility, and the effects of higher altitudes. Leveraging machine learning approaches can help address the challenges posed by this complex system model.}
\bibliography
{References/References}
\end{document}